\documentclass[a4paper,fleqn]{cas-sc}

\usepackage[authoryear,round]{natbib}
\usepackage{xcolor}
\usepackage[normalem]{ulem}

\def\tsc#1{\csdef{#1}{\textsc{\lowercase{#1}}\xspace}}
\tsc{WGM}
\tsc{QE}
\tsc{EP}
\tsc{PMS}
\tsc{BEC}
\tsc{DE}
\usepackage{lineno}

 \usepackage{amsmath,amssymb,amsfonts}    
 \usepackage{graphicx}                    
 \usepackage{algorithm}                   
 \usepackage{algorithmic}                 
\usepackage{listings}
\usepackage{xcolor}
\definecolor{revision}{RGB}{0,0,255}
\lstdefinelanguage{Julia}{
  morekeywords={
    abstract,break,case,catch,const,continue,do,else,elseif,end,
    export,false,for,function,immutable,import,importall,if,in,
    isa,let,macro,module,quote,return,true,try,type,typealias,
    using,while
  },
  sensitive=true,
  morecomment=[l]\#,
  morestring=[b]"
}

\usepackage{float}
\usepackage{placeins} 
\usepackage{hyperref}
\hypersetup{
    colorlinks=true,       
    linkcolor=blue,        
    citecolor=blue,        
    filecolor=magenta,     
    urlcolor=blue          
}
\usepackage{caption}
\begin{document}
\let\WriteBookmarks\relax
\def\floatpagepagefraction{1}
\def\textpagefraction{.001}
\shorttitle{Backend-agnostic Julia framework for 3D modeling and inversion of gravity data }

\shortauthors{Nimatullah et al.}

\title [mode = title]{\textbf{Backend-agnostic Julia framework for 3D modeling and inversion of gravity data}}

\author[1]{Nimatullah}
\author[2]{Pankaj K Mishra} 
\author[2]{Jochen Kamm}
\author[1]{Anand Singh}

\address[1]{Department of Earth Sciences, Indian Institute of Technology,
Bombay, India}
\address[2]{Geological Survey of Finland, Vuorimiehentie 5, Espoo, Finland}

\begin{abstract}
\normalsize \noindent 

This paper presents a high-performance framework for three-dimensional gravity modeling and inversion implemented in Julia, addressing key computational challenges associated with gravity inversion, including large scale problem, ill-posedness, and non-uniqueness. 
The framework employs a data-space inversion formulation that reduces the dimensionality of the inverse problem, resulting in lower memory requirements and improved computational efficiency.
Forward modeling and inversion operators are implemented using a backend-agnostic kernel abstraction, allowing the same computational code to run on multicore CPUs and GPU accelerators. 
Performance evaluations using NVIDIA CUDA GPUs show significant reductions in computational time relative to CPU execution, particularly for large-scale problems involving up to approximately 3.3 million rectangular prisms and 178,797 gravity observations. 
The inversion incorporates implicit model constraints through the data-space formulation and depth-weighted sensitivity, which reduces the effect of depth-dependent amplitude decay and promotes geologically coherent density models. 
Synthetic experiments demonstrate the capability of the framework to recover complex subsurface structures, including vertical and dipping dykes.
Application to field gravity data further demonstrates the practical applicability of the proposed approach, with the recovered density distributions showing good agreement with independent geological constraints and previous interpretations. 
The results demonstrate that GPU-accelerated Julia provides an efficient and extensible platform for large-scale three-dimensional gravity modeling and inversion, enabling high-resolution geophysical investigations with reduced computational and memory requirements.

\end{abstract}





\maketitle

\setlength{\parindent}{0pt}
\setlength{\parskip}{5pt} 
\normalsize 
\section{Introduction} 
\linespread{1.15}
Three-dimensional (3D) gravity inversion is a vital geophysical technique to infer subsurface density distributions from gravity anomaly data. This method is widely applied in resource exploration, geodynamic studies, and environmental investigations to model geological structures such as ore bodies, sedimentary basins, and crustal heterogeneities \citep{tirel2004aegean, jacoby2009gravity, reynolds2011applied, singh2019triangular} and is important first-step towards a lithospheric digital twin \citep{Mishra2025DT}. By converting observed gravity anomalies into a 3D density distribution, inversion techniques provide critical insights into subsurface properties without requiring direct drilling or intrusive measurements. However, 3D gravity inversion is inherently non-unique, as gravity observations alone are insufficient to uniquely determine the underlying density distribution. Multiple subsurface models can reproduce the same gravity response. Furthermore, the presence of noise, limited data coverage, and model discretization renders the inversion problem ill-posed, necessitating the use of regularization to obtain stable and physically meaningful solutions.
\citep{li1998inversion, camacho2000gravity, camacho2002gravity, rim2007euler, gottsmann2008shallow}. Researchers have developed various inversion algorithms to address this issue, often incorporating regularization techniques to stabilize the solution and reduce ambiguities. One of the most widely used methods is L$_2$-norm regularization, which enforces smoothness constraints to prevent unrealistic density variations in the solution \citep{oldenburg1994subspace, li1998inversion, hightower2020bayesian}.

Despite its effectiveness, L$_2$-norm regularization has notable limitations. It tends to smooth out sharp density contrasts, which can obscure critical geological features such as ore bodies, fault zones, or lithological boundaries. To overcome this, alternative approaches have been explored, including L$_1$ norm and sparsity-promoting regularization techniques, which improve resolution by preserving sharp edges in the reconstructed density model \citep{jafarpour2009transform, utsugi2019magnetic, meng2018three}. While such approaches can enhance edge preservation, the present study focuses on a computationally efficient quadratic data space inversion framework, emphasizing scalability and performance rather than sparsity-promoting regularization.

In addition, joint inversion algorithms that integrate gravity data with complementary geophysical datasets, such as seismic, magnetic, or electrical resistivity data, have been developed to improve resolution and reduce uncertainty in subsurface models \citep{carrillo2018joint, lelievre2012joint, moorkamp2011framework, fregoso2009cross}. Incorporating geological constraints, such as well log data, lithological boundaries, and previous geological knowledge, further refines the results of inversion and improves interpretability \citep{bosch2001joint, calcagno2008geological,kamm2015joint}.

Another major challenge in 3D gravity inversion is computational complexity, especially when dealing with large-scale datasets. Traditional CPU-based implementations can become computationally prohibitive in terms of runtime and memory consumption, as the number of model unknowns increases.

In response to these computational challenges, this study presents a high-performance 3D gravity inversion framework implemented in Julia, a modern programming language designed for high-performance scientific computing \citep{bezanson2017julia, jones2020seisio}. The framework leverages data space inversion to reduce the dimensionality of the problem, thereby improving computational efficiency while maintaining inversion accuracy \citep{rezaie2017gravity, marchetti2014large}.

The data space approach used in this study builds on previous
developments in large-scale geophysical inversion that reformulate the
inverse problem in the observation domain to avoid the direct solution
of the full model-space system. Such formulations have been developed
for large-scale inverse problems and applied to magnetic,
electromagnetic, magnetotelluric, and potential-field inversion
\citep{oldenburg1994subspace,pilkington2009magnetic,
siripunvaraporn2005dataspace,vatankhah2022gramian}. In the present
study, we combine this data space formulation with an analytical
closed-form prism forward operator and a backend-agnostic GPU kernel
framework, allowing the same computational framework to support CPU,
single-GPU, and multi-GPU execution. This combination enables
large-scale 3D gravity inversion while reducing the computational
and memory requirements associated with the model-space formulation,
and differs from previous GPU-accelerated potential-field inversion
approaches such as \citep{cuma2014gpu}, which did not employ a
data space reduction.

The framework is designed to support both CPU and vendor-neutral GPU execution. GPU acceleration is implemented and benchmarked on NVIDIA CUDA GPUs, significantly enhances computational performance by parallelizing large-scale linear algebra operations and reducing data transfer bottlenecks. Benchmarking results show that GPU-based execution leads to substantial reductions in runtime compared to CPU-based computations, especially for large-scale problems involving models with up to approximately 3.3 million rectangular prisms. The framework retains numerical accuracy and stability across both architectures, making it suitable for large-scale inversion using standard as well as physics-based machine learning method \citep{mishra2025inr} through accelerated and differentiable forward solver.

Three real-data applications are considered to demonstrate complementary
capabilities of the proposed framework. The Tangarparha, Odisha, dataset
represents an ore-exploration application, where the inversion is used
to recover the location, geometry, and depth extent of dense
mafic/ultramafic bodies associated with chromite mineralization. The
Bharatpur, Rajasthan, dataset is used to investigate subsurface density
variations associated with BIF and other geological formations and their
relationship with the regional geological framework. In addition, a
large-scale gravity dataset from Bhukia, Rajasthan, is used to evaluate
the computational scalability of the proposed framework for a
million-cell regional inversion. These applications were selected to
provide complementary tests of the method, ranging from localized
ore-body imaging and geological interpretation to large-scale
computational validation.

This paper evaluates the proposed GPU-accelerated 3D gravity inversion
framework through a series of synthetic and field-based examples. The
synthetic tests confirm the method’s ability to resolve complex
subsurface structures such as dipping and vertical dykes, while the

field applications demonstrate complementary capabilities. These include
localized ore-body geometry recovery at Tangarparha, subsurface density
and geological-pattern interpretation at Bharatpur, and large-scale
computational validation at Bhukia. Together, these examples evaluate
both the geological applicability and computational scalability of the
proposed framework.

\section{Methodology}

\subsection{3D Gravity Forward Modeling}

The gravitational response of subsurface structures is computed using a vertex-centered approach. For a rectangular prism defined by vertices $(x_p, y_q, z_r)$ where $p, q, r \in {1, 2}$, the vertical gravity component at the observation point $(0, 0)$ is given by \citep{blakely1996potential}:

\begin{equation}
\delta g_z^k =\gamma \rho^k \sum_{p=1}^{2} \sum_{q=1}^{2} \sum_{r=1}^{2} \sigma_{pqr} \Psi(x_p, y_q, z_r)
\end{equation}

where $\sigma_{pqr} = (-1)^{p + q + r}$ is the vertex parity operator, and $\Psi(x_p, y_q, z_r)$ is defined as:

\begin{equation}
\Psi(x_p,y_q,z_r)
=
z_r \tan^{-1}\!\left(
\frac{x_p y_q}{z_r R_{pqr}}
\right)
- x_p \ln\!\left({R_{pqr}}+y_q\right)
- y_q \ln\!\left({R_{pqr}}+x_p\right),
\end{equation}

with:

\begin{equation}
R_{pqr} = \sqrt{x_p^2 + y_q^2 + z_r^2}.
\end{equation}

Here, $\gamma$ denotes the gravitational constant and $R_{pqr}$ is the distance from the observation point to the $(p,q,r)$-th prism vertex.

The total gravity anomaly due to all cells can be computed as the sum 
of the gravity anomalies produced by the individual cells at the 
observation point $(0,0)$. Thus,

\begin{equation}
g_z = \sum_{k=1}^{M} \delta g_z^k,
\end{equation}

where $M$ is the total number of cells. Let $(x_0^i,y_0^i)$, $i=1,2,3,\ldots,N$, denote the coordinates of the data points, where $N$ is the total number of data points in the survey  area. The vertical component of the gravitational field at the $i$th 
data point can then be expressed as

\begin{equation}
d_i = \sum_{k=1}^{M} g_z^{i,k}\rho^k,
\end{equation}

where $\rho^k$ is the density of the $k$th cell and 
$g_z^{i,k}$ is the gravity response coefficient representing the influence 
of the $k$th cell on the $i$th data point. The coefficient $g_z^{i,k}$ 
can be formulated using Eqs.~(1) and (2).

Equation~(5) can be formulated in matrix form as

\begin{equation}
\mathbf{d}^{\mathrm{cal}} = \mathbf{G}\mathbf{m}.
\end{equation}

\subsection{3D Gravity Inversion}

The subsurface density distribution is estimated by minimizing the objective function:
\begin{equation}
\min_{\mathbf{m}} \phi(\mathbf{m}) =
(\mathbf{d}^{\text{obs}} - {\mathbf{G}}\mathbf{m})^\top \mathbf{W}_d^{-1} (\mathbf{d}^{\text{obs}} - {\mathbf{G}}\mathbf{m}) +
(\mathbf{m} - \mathbf{m}_{\text{ref}})^\top \mathbf{W}_m^{-1} (\mathbf{m} - \mathbf{m}_{\text{ref}})
\end{equation}

In this formulation, $\mathbf{d}^{\text{obs}}$ is the observed gravity anomaly vector ($N \times 1$), ${\mathbf{G}}$ is the forward modeling matrix ($N \times M$), $\mathbf{m}$ is the density parameter vector ($M \times 1$), $\mathbf{m}_{\text{ref}}$ is the reference model ($M \times 1$), $\mathbf{W}_d$ and $\mathbf{W}_m$ are the data weighting matrix and regularization matrix ($N \times N$ and $M \times M$, respectively).

The data weighting is represented by a diagonal data space covariance term with a uniform value $\delta$,
whereas the model regularization is represented by a diagonal
depth-weighting term. The depth weighting compensates for the decrease
in gravity sensitivity with depth and reduces the tendency of the inversion
to concentrate density variations near the observation surface, following
the depth-weighting approach of \cite{li1998inversion}. Thus, the adopted
regularization is quadratic and depth-weighted rather than explicitly
sparsity-promoting.

The reference model $\mathbf{m}_{\mathrm{ref}}$ is initialized as a uniform
density-contrast model with a reference value of
$0.0001~\mathrm{g/cm^3}$.
No prior ensemble is generated, and geological information is not
directly imposed as an inversion constraint. This quadratic depth-weighted regularization also introduces some
smoothing of the recovered density distribution, particularly near
sharp anomaly boundaries and at greater depths.

Differentiating with respect to $\mathbf{m}$ yields the linear system:
\begin{equation}
\left( {\mathbf{G}}^\top \mathbf{W}_d^{-1} {\mathbf{G}} + \mathbf{W}_m^{-1} \right) (\mathbf{m} - \mathbf{m}_{\text{ref}}) =
{\mathbf{G}}^\top \mathbf{W}_d^{-1} (\mathbf{d}^{\text{obs}} - {\mathbf{G}} \mathbf{m}_{\text{ref}})
\label{eq:direct}
\end{equation}

Direct inversion of (\ref{eq:direct}) is computationally prohibitive for $M \sim 10^5$--$10^6$.

Using the matrix identity \citep{wang2015computationally}:
\begin{equation}
({\mathbf{G}}^\top \mathbf{W}_d^{-1} {\mathbf{G}} + \mathbf{W}_m^{-1})^{-1} {\mathbf{G}}^\top \mathbf{W}_d^{-1} =
\mathbf{W}_m {\mathbf{G}}^\top ({\mathbf{G}} \mathbf{W}_m {\mathbf{G}}^\top +\mathbf{W}_d)^{-1}
\end{equation}

the model update can be written as:

\begin{equation}
\delta \boldsymbol{m}  = \mathbf{W}_m {\mathbf{G}}^{T}({\mathbf{G}} \mathbf{W}_m {\mathbf{G}}^{T} + \mathbf{W}_d)^{-1}(\mathbf{d}^{\text{obs}}-{\mathbf{G}}\boldsymbol{m}_{\text{ref}})
\end{equation}

where $\delta \boldsymbol{d} = \boldsymbol{d}^{obs}-{\mathbf{G}}\boldsymbol{m}_{ref}$.

The solution can therefore be written as:

\begin{equation}
\delta \mathbf{m} = \mathbf{W}_m {\mathbf{G}}^\top \left( {\mathbf{G}} \mathbf{W}_m {\mathbf{G}}^\top + \mathbf{W}_d \right)^{-1} \delta \mathbf{d}
\end{equation}

where $\delta \mathbf{m} = \mathbf{m} - \mathbf{m}_{\text{ref}}$ and $\delta \mathbf{d} = \mathbf{d}^{\text{obs}} - {\mathbf{G}} \mathbf{m}_{\text{ref}}$. This reduces the system size to $N \times N$ ($N \ll M$).

This reformulation also reduces the memory required for the inversion
because the dense system to be solved is reduced from $M\times M$ to
$N\times N$, where $N\ll M$. In the present work, data space inversion refers to the algebraic
reformulation of the quadratic gravity inverse problem in the data
domain rather than the model domain. Unlike ensemble-based approaches,
the present method does not generate a prior ensemble of 3D density
models, forward model each realization, or estimate model--data
covariance statistically. Instead, the sensitivity matrix $\mathbf{G}$,
together with the data weighting matrix $\mathbf{W}_d$ and model
regularization matrix $\mathbf{W}_m$, is used directly to construct the
$N\times N$ data space system
$\mathbf{G}\mathbf{W}_m\mathbf{G}^{T}+\mathbf{W}_d$. The discrepancy
between the observed data and the response of the reference model is
defined as
$\delta\mathbf{d}=\mathbf{d}^{\mathrm{obs}}
-\mathbf{G}\mathbf{m}_{\mathrm{ref}}$. The resulting data space system
is then solved to obtain the model update $\delta\mathbf{m}$ through
the data space formulation given above.

For each inversion experiment, the procedure produces a single deterministic density model through iterative model
updates rather than an ensemble of models. For the reported inversions, the algorithm was configured with a maximum
of 30 outer model-update iterations, with up to 20 inner conjugate-gradient iterations per outer iteration. The outer
iteration terminates when the residual criterion is satisfied ($\mathrm{res} \leq 10^{-4}$) or when the maximum of 30
outer iterations is reached. The inner conjugate-gradient solve is similarly limited to 20 iterations and uses the same
tolerance. Thus, the reported solution is a single deterministic model obtained through the specified iterative
procedure; no prior or posterior model ensemble is generated. Because the present formulation does not generate an
ensemble, posterior variance and probability maps are not directly computed. Formal resolution analysis and
probabilistic uncertainty quantification are beyond the scope of the present study.

To further accelerate computations, especially for large-scale problems, we compare the performance of our inversion code executed on a central processing unit (CPU) and a graphics processing unit (GPU). Both implementations utilize the same numerical algorithm and data but differ in computational architecture. The GPU version leverages parallel processing to perform matrix operations, particularly matrix multiplications and factorizations, more efficiently than the CPU.

The GPU acceleration methodology employs a multifaceted parallelization strategy to overcome computational bottlenecks in 3D gravity inversion. Backend-agnostic GPU kernels, executed on the CUDA backend, distribute gravitational response calculations across thousands of GPU threads, with each thread handling specific observation-cell pairs using singularity-robust formulations. Optimization of the memory hierarchy strategically places data across global memory (sensitivity matrix), shared memory (cell parameters), constant memory (physical constants), and register storage (intermediate results) to maximize throughput.

The GPU kernels are implemented using the backend-agnostic
KernelAbstractions.jl interface, allowing the same kernel
implementation to target supported CPU and GPU backends. In the
present study, however, all GPU performance experiments were conducted
using the NVIDIA CUDA backend; other GPU backends were not
experimentally evaluated.

The real-data applications considered in this study use ground-based
gravity observations. The gravity measurements were processed to obtain
Bouguer gravity anomalies, which contain contributions from both
regional and local subsurface density variations. To emphasize the
density variations relevant to the 3D inversion, the regional component was separated from the Bouguer anomaly using a second-degree polynomial trend-surface fit. The resulting residual Bouguer gravity anomaly was
then used as the input for the 3D gravity inversion. All real-data applications presented in this study are based on
ground-based (land) gravity observations; no satellite- or
airborne-derived gravity data are used in the present study.

\section{Computation and Validation}
\subsection{Backend-agnostic kernel abstraction} 
The parallel implementation in this work is based on KernelAbstractions.jl, which provides a hardware-agnostic abstraction for writing GPU-style kernels in Julia. Kernels are defined once using an abstract execution model and can be executed on different compute backends by changing only the backend selection at runtime. As a result, the same kernel code can run on multi-core CPUs or GPUs without modification. This approach differs from conventional GPU acceleration strategies, where separate device-specific forward kernels are typically required for CPU and GPU execution. In KernelAbstractions.jl, backend-specific details such as thread organization, indexing, and memory execution are handled internally, allowing the kernel logic to remain independent of the target hardware. Kernels developed and validated on the CPU can therefore be executed directly on the GPU using the same source code. In addition to CPU--GPU portability, kernel abstraction enables cross-vendor GPU portability. The same kernel formulation can, in principle, be executed on different GPU architectures, including NVIDIA (CUDA), AMD, and Apple Metal backends.

Although the performance results in this study are reported only for the
CUDA backend, the kernel implementations are designed using the
backend-agnostic \texttt{KernelAbstractions.jl} interface. AMDGPU, Metal,
and oneAPI backends were not experimentally benchmarked in this study.
Therefore, the reported performance results are limited to the CUDA
implementation. An illustrative kernel implementation is provided in
Listing~\ref{lst:kernelabstraction}.

\begin{lstlisting}[language=Julia,
caption={Illustrative backend-agnostic kernel implementation using
\texttt{KernelAbstractions.jl}. The same kernel definition can be
dispatched to supported computational backends. Only the NVIDIA CUDA
backend was experimentally evaluated in this study.},
label={lst:kernelabstraction}]
@kernel function gravity_kernel!(G, ...)
    i = @index(Global)
    if i <= length(G)
        G[i] = compute_gravity(...)
    end
end

backend = CUDA.CUDABackend()
gravity_kernel!(backend, ...; ndrange=length(G))
\end{lstlisting}

\subsection{Software Architecture and Design Principles}
\label{sec:multigpu-method}
\begin{figure}[htbp]
    \centering
    \includegraphics[width=\textwidth]{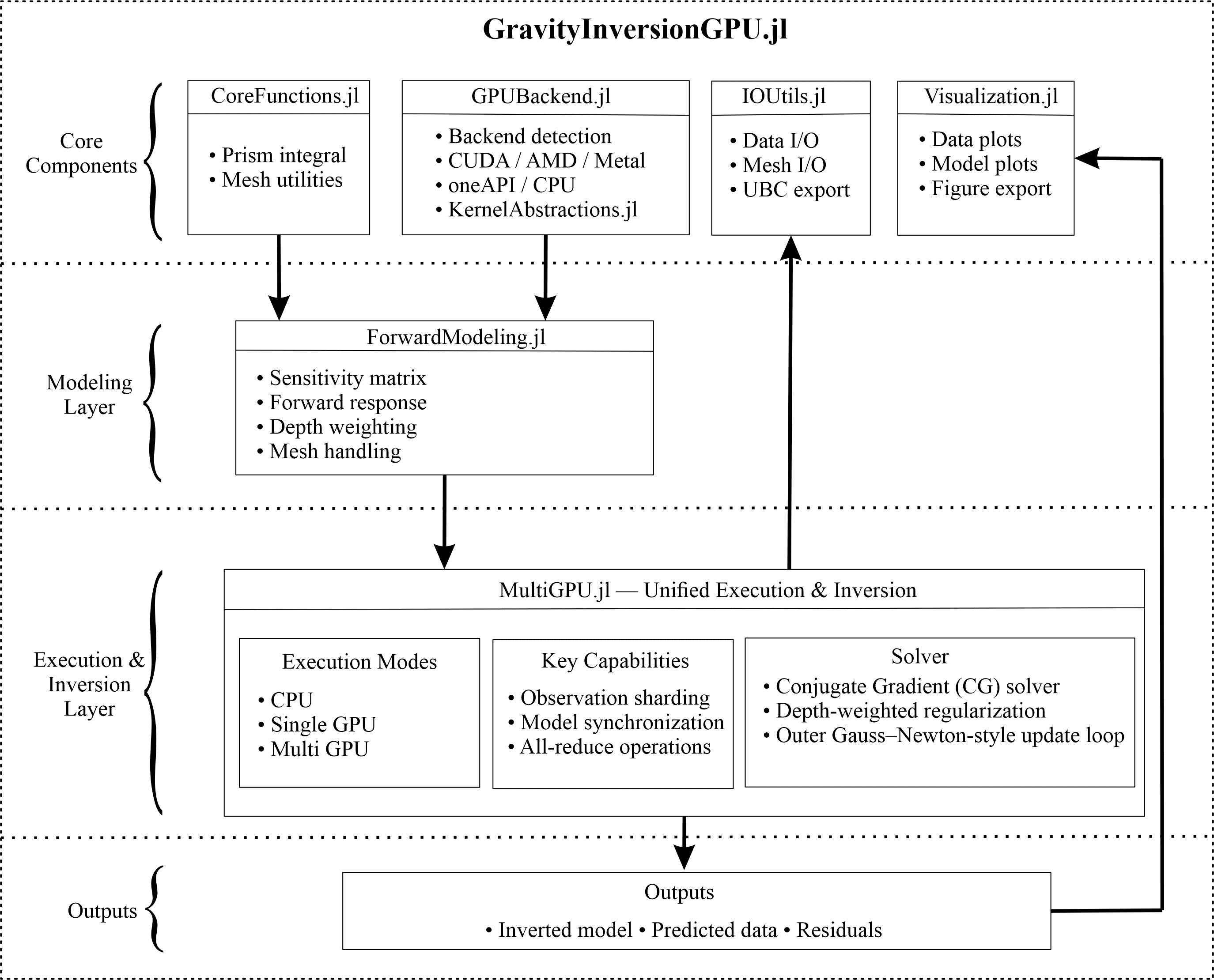}
    \caption{Software architecture of the \texttt{GravityInversionGPU.jl} framework. The framework is organized into core components, a forward-modeling layer, a unified execution and inversion layer, and an output layer. The unified execution engine supports CPU, single-GPU, and multi-GPU execution through a common implementation.}
    \label{fig:software_architecture}
\end{figure}
The framework follows modular software engineering principles rather than being implemented as a single monolithic program. As shown in Fig.~\ref{fig:software_architecture}, the package is organized into four functional layers: core components, modeling, execution and inversion, and outputs. Each component has a well-defined responsibility and communicates with the others through clearly defined interfaces, an organization that supports modularity, separation of concerns, low coupling, and long-term maintainability.

The core components provide the reusable functionality on which the rest of the framework depends. \texttt{CoreFunctions.jl} contains general-purpose mathematical and grid-related utilities, including the analytical prism integral and mesh operations. \texttt{GPUBackend.jl} isolates hardware management from the numerical algorithms, providing automatic backend detection and a unified interface through \texttt{KernelAbstractions.jl}; this abstraction lets the same computational kernels run on CPU, CUDA, AMDGPU, Metal, or oneAPI without hardware-specific changes elsewhere in the code. Input/output and visualization are similarly kept separate, in \texttt{IOUtils.jl} and \texttt{Visualization.jl} respectively, so that file handling and plotting never become entangled with the core computational routines.

The modeling layer, implemented in \texttt{ForwardModeling.jl}, constructs the gravity sensitivity matrix, computes the forward response, applies depth weighting, and manages the associated mesh and data structures. Keeping forward modeling separate from execution and inversion means the forward problem can be developed independently of how its computation is later distributed across hardware; the resulting operator is then passed to the execution and inversion layer through a well-defined interface.

The central computational component, \texttt{MultiGPU.jl}, combines device-independent execution and inversion within a single implementation. The same engine supports CPU, single-GPU, and multi-GPU configurations: depending on the resources available, the observation domain is partitioned across devices while model-space variables are replicated and synchronized as the inversion algorithm requires. 

Within \texttt{MultiGPU.jl}, inversion is carried out by a conjugate-gradient (CG) solver with depth-weighted regularization. An outer Gauss--Newton-style update loop with step-size backtracking governs the iterative model updates. Keeping the execution strategy and the solver in the same module allows sharding, synchronization, and iterative inversion to be coordinated directly, without maintaining separate hardware-specific solver implementations.

Results from the execution and inversion layer feed into the output stage, comprising the inverted model, predicted data, residuals, and derived figures. These outputs are then consumed by \texttt{IOUtils.jl} for storage and by \texttt{Visualization.jl} for graphical representation, a one-way flow that keeps the numerical computation independent of data export and visualization.

This modular organization also makes the framework easier to extend. New forward-modeling capabilities, inversion strategies, regularization schemes, visualization routines, or hardware backends can be added by extending the relevant component alone, without restructuring the rest of the framework. The backend abstraction in particular separates hardware-specific execution from the numerical kernels, so additional backends can be supported without rewriting the core algorithms.

Overall, the architecture favors high cohesion within each component and low coupling between them, with all interaction occurring through well-defined interfaces. This reduces code duplication, simplifies maintenance and debugging, and gives the framework a consistent computational interface as it grows to support additional large-scale geophysical modeling and inversion applications.

\subsection{Computational Performance Comparison in Gravity Inversion}

\begin{figure}[h!]
  \centering
  \includegraphics[width=1\textwidth]{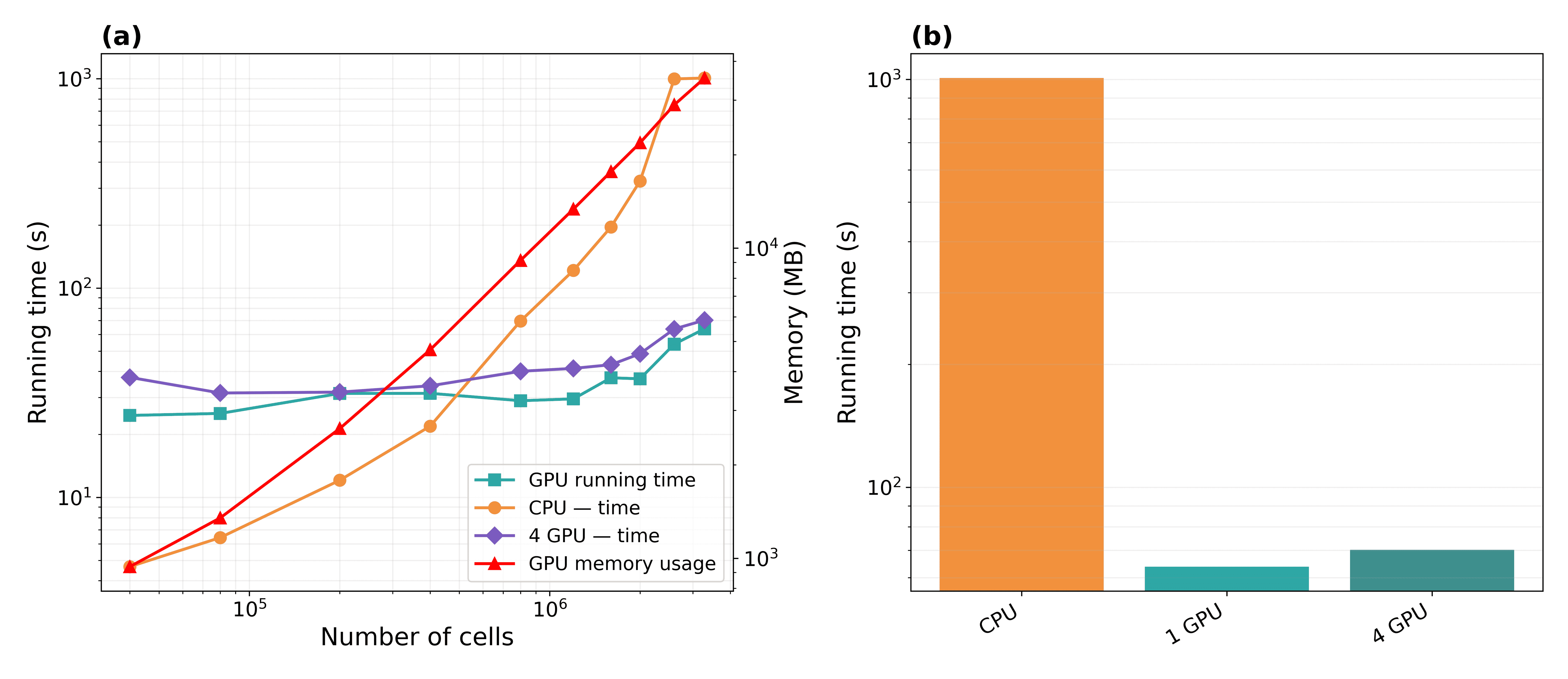} 
\caption{Computational performance and memory scaling of the proposed
framework. (a) Total computational runtime for CPU, single-GPU, and
four-GPU configurations, together with GPU memory usage, as a function
of the number of cells. (b) Total runtime comparison for the largest
$3.28\times10^{6}$-cell problem. All GPU calculations use Float32
precision on NVIDIA A100 GPUs.}
  \label{fig:Figure1}
\end{figure}

Figure~\ref{fig:Figure1} summarizes the computational performance and
GPU memory usage across the model-size sweep. The CPU runtime increases
rapidly with model size, whereas GPU execution scales more favorably.
The four-GPU configuration provides additional acceleration for the
largest problems, although communication and synchronization overheads
limit the ideal scaling. GPU memory usage increases approximately
linearly with the number of cells, reaching about 34.1~GB for the
$3.28\times10^{6}$-cell problem.

As the model size increases, the computational characteristics change fundamentally. Beyond approximately $10^{3}$--$10^{4}$ rectangular prisms, the arithmetic workload becomes sufficiently large for the GPU to exploit its massively parallel architecture. The CPU runtime increases rapidly with problem size, while the GPU runtime increases much more slowly with problem size and remains relatively stable over several orders of magnitude. For large-scale models exceeding $10^{5}$ cells and extending up to about $3.3\times10^{6}$ rectangular prisms, CPU computation times surpass $10^{3}$ seconds, whereas the GPU performs the same computation in approximately $20$--$30$ seconds, yielding one to two orders of magnitude speedup.
For problem sizes larger than approximately $3.3\times10^{6}$ prisms, the computation could not be completed on the Mahti GPU system due to a memory error.

\begin{table}[htbp]
\centering
\caption{Measured total wall-clock time and GPU memory usage for
the controlled 2,601-observation benchmark. Total time includes
forward-model construction, inversion, and associated computational
steps.}
\label{tab:benchmark_total}
\begin{tabular}{@{}lcc@{}}
\toprule
 & $2.0\times10^6$ cells & $3.28\times10^6$ cells \\
\midrule
CPU total time & 324.21 s & 1007.46 s \\
1$\times$ A100 total time & 36.87 s & 63.88 s \\
4$\times$ A100 total time & 48.57 s & 70.27 s \\
CPU $\rightarrow$ 1 GPU speedup & 8.79$\times$ & 15.77$\times$ \\
CPU $\rightarrow$ 4 GPU speedup & 6.67$\times$ & 14.33$\times$ \\
Peak GPU memory, 1$\times$ A100 & 21.85 GB & 35.34 GB \\
Peak GPU memory, 4$\times$ A100 & 6.21 GB/device & 10.44 GB/device \\
\bottomrule
\end{tabular}
\end{table}

\begin{table}[t]
\centering
\caption{Measured wall-clock times and GPU memory usage for representative
synthetic and field-data inversions using CPU and GPU implementations.}
\label{tab:actual_runtime}
\begin{tabular}{@{}lrrrrrrr@{}}
\toprule
\textbf{Dataset} & \textbf{Type} & \textbf{Cells} & \textbf{Obs.}
& \textbf{CPU (s)} & \textbf{1 GPU (s)} & \textbf{4 GPU (s)}
& \textbf{GPU Memory (GB)} \\
\midrule
Two bodies & Synthetic & 2.0M & 4,356
& 451.92 & 59.06 & 63.13 & 35.71 / 9.63 \\

Curved tube + dipping slab & Synthetic & 0.5M & 38,416
& 1928.18 & -- & 92.63 & -- / 19.83 \\

Tangarparha & Field & 2.81M & 891
& 70.66 & 59.34 & 90.16 & 10.02 / 2.50 \\

Bharatpur & Field & 1.34M & 6,469
& 194.98 & 108.29 & 102.85 & 34.66 / 8.67 \\
\bottomrule
\end{tabular}

\begin{flushleft}
\footnotesize
GPU Memory values are reported as 1-GPU peak memory / 4-GPU peak
memory per device. A dash indicates that the configuration was not
feasible because the dense Float32 sensitivity matrix for the curved
synthetic model was 76.832~GB, exceeding the 40~GB memory capacity of a
single NVIDIA A100 GPU.
\end{flushleft}
\end{table}

Measured runtimes for representative synthetic and field inversions are
summarized in Table~\ref{tab:actual_runtime}. The GPU implementation
reduces the computational time relative to the CPU for the synthetic
examples and the Bharatpur field case. For the smaller problems,
multi-GPU execution provides limited additional benefit because of
parallelization and communication overhead.

All benchmark calculations were performed on the CSC Mahti supercomputer
using Float32 precision and NVIDIA A100 GPUs with 40~GB memory per GPU.
The CPU baseline used an AMD EPYC Rome 7H12 processor with 16 CPU threads
for sensitivity-matrix construction, with BLAS threads capped at 8 during
the iterative inversion. The sensitivity-matrix construction and inversion
stages were timed separately. Data-transfer operations occurring within
these computational stages are included in the reported timings, while
subsequent data export and visualization are excluded. The inversion uses
an iterative conjugate-gradient solver and does not require explicit matrix
factorization. For the largest problem, the dense Float32 sensitivity
matrix was 34.13~GB, with a measured peak GPU memory usage of 35.34~GB
on one A100 and 10.44~GB per GPU on four A100 GPUs.

\FloatBarrier

\section{Example Applications}

\subsection{Synthetic Example – Two-Body Model}

The first synthetic experiment was designed to evaluate the capability of the proposed data space inversion framework to recover multiple isolated density anomalies with different lateral positions and depth extents. Two rectangular prismatic bodies embedded in a homogeneous background were considered to provide a controlled validation scenario.


The true density model consists of two rectangular prismatic bodies
embedded in a homogeneous background. Both bodies were assigned a density
contrast of $+1.0$~g/cm$^3$. The first body extends from approximately
13,000~m to 17,000~m in both the \textit{x}- and \textit{y}-directions
and from 1,500~m to 3,000~m in depth. The second body extends from
approximately 6,000~m to 10,000~m in both the \textit{x}- and
\textit{y}-directions and from 500~m to 3,000~m in depth. This
configuration provides two spatially separated anomalies at different
depth ranges for evaluating the ability of the inversion to recover
multiple subsurface bodies.

Synthetic gravity data were first generated from the true model using the
forward modeling operator. To assess the robustness of the proposed
inversion to observational noise, additive white Gaussian noise was
introduced into the synthetic gravity data. Noise levels of 0\%, 1\%, and
2\% RMS noise were considered. The resulting datasets were independently
inverted using the data-space inversion approach without imposing
a priori constraints on the geometry or number of anomalous bodies.

\begin{figure}[h!]
  \centering
  \includegraphics[width=1\textwidth]{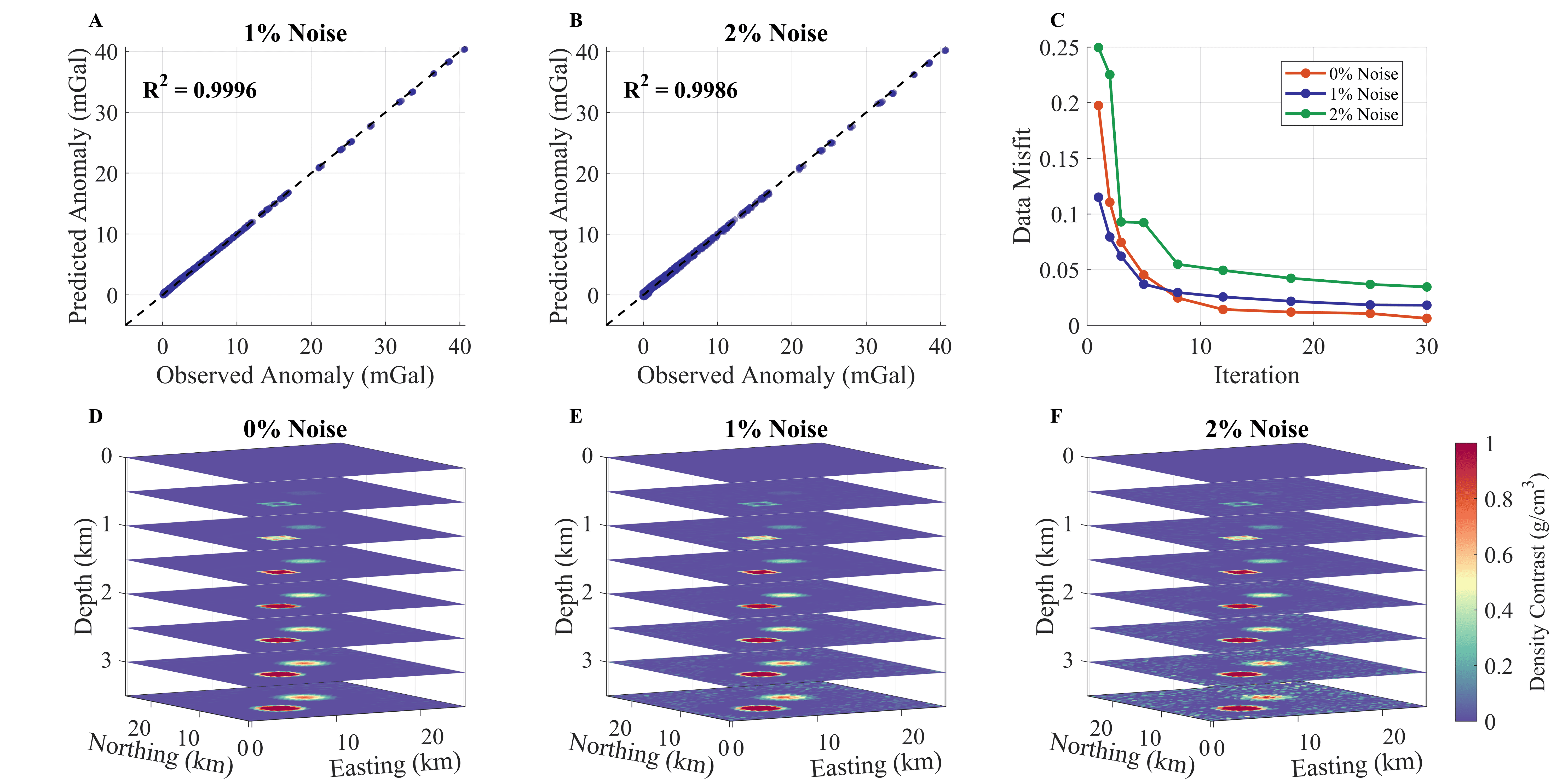}
  \caption{
  Noise-sensitivity analysis of the synthetic two-body gravity inversion.
  Panels A and B compare the observed and calculated gravity responses
  for 1\% and 2\% white Gaussian noise, respectively. The dashed line
  represents the one-to-one relationship, with coefficients of determination
  of $R^2=0.9996$ and $R^2=0.9986$, respectively. Panel C shows the
  convergence of the relative data misfit with iteration for the 0\%, 1\%,
  and 2\% noise cases. Panels D--F present the recovered
  density-contrast distributions at successive depth levels for 0\%, 1\%,
  and 2\% noise, respectively. Increasing noise introduces greater
  variability in the recovered structures, particularly at greater depths,
  while the principal subsurface structures remain identifiable.
  }
  \label{fig:noise_sensitivity}
\end{figure}

Figure~\ref{fig:noise_sensitivity} summarizes the effect of increasing
observational noise on the gravity response, convergence, and recovered
density structure. The top row (Figures~\ref{fig:noise_sensitivity}A--B)
shows the observed-versus-calculated gravity responses for the 1\% and
2\% noise cases. The calculated responses remain closely aligned with
the one-to-one relationship, with $R^2=0.9996$ for 1\% noise and
$R^2=0.9986$ for 2\% noise, indicating that the dominant gravity signal
is well reproduced even in the presence of noise.

The middle panel (Figure~\ref{fig:noise_sensitivity}C) compares the
convergence of the data misfit for the three noise levels. In all cases,
the misfit decreases rapidly during the initial iterations and then
approaches a more gradual convergence. The noise-free case reaches the
lowest final misfit, while the 1\% and 2\% noise cases retain higher
misfit values because of the additional variability introduced into the
observations.

The bottom row (Figures~\ref{fig:noise_sensitivity}D--F) presents the
recovered density-contrast distributions at successive depth levels.
The noise-free result shows well-defined and spatially coherent
anomalies. With 1\% noise, the two principal bodies remain clearly
identifiable, although some additional variability is visible at greater
depths. At 2\% noise, the recovered structures show greater spatial
variability and reduced continuity, particularly at greater depths. Nevertheless, the locations and overall
geometry of the two principal bodies remain recognizable.

Overall, the results demonstrate that increasing relative observational
noise progressively affects the data fit and the resolution of the
recovered density model. The effect is more pronounced at greater depths,
where gravity sensitivity to subsurface structures is reduced. Despite
this degradation, the principal anomalous bodies remain identifiable at
2\% noise, while the high $R^2$ values demonstrate that the proposed
data-space inversion framework remains robust to the tested levels of
relative Gaussian noise.

\subsection{Complex Synthetic Example – Curvilinear and Dipping Bodies}

To further evaluate the capability of the proposed data space inversion
framework for geometrically complex subsurface structures, a second
synthetic model containing two non-axis-aligned bodies was considered.
Unlike the rectangular bodies used in the previous synthetic experiment,
this model consists of a curvilinear tube with spatially varying width and
a dipping slab, providing a more challenging test of the inversion
framework.

The first body is a sinuous tube whose centerline follows a curved path in
the horizontal plane. Its width varies along the centerline, producing
alternating broadened and narrowed portions, and it extends from
1{,}000~m to 4{,}000~m depth. The second body is a tabular dipping slab
with a $40^\circ$ dip, extending along the easting direction and becoming
progressively deeper toward increasing northing. The two bodies are
spatially separated and do not overlap. The curvilinear tube was assigned
a density contrast of $-1.0$~g/cm$^3$, whereas the dipping slab was assigned
a density contrast of $+1.0$~g/cm$^3$ relative to the zero-density
background. The complex geometry was discretized on a
$100\times100\times50$ mesh with cell dimensions of
$200\times200\times100$~m.

Synthetic gravity observations were generated by forward modeling the
complex true model using the same forward operator employed in the
inversion. The resulting synthetic data were subsequently inverted using
the proposed data space inversion approach without imposing a priori
constraints on the geometry or number of anomalous bodies.

\begin{figure}[h!]
\centering
\includegraphics[width=1\textwidth]{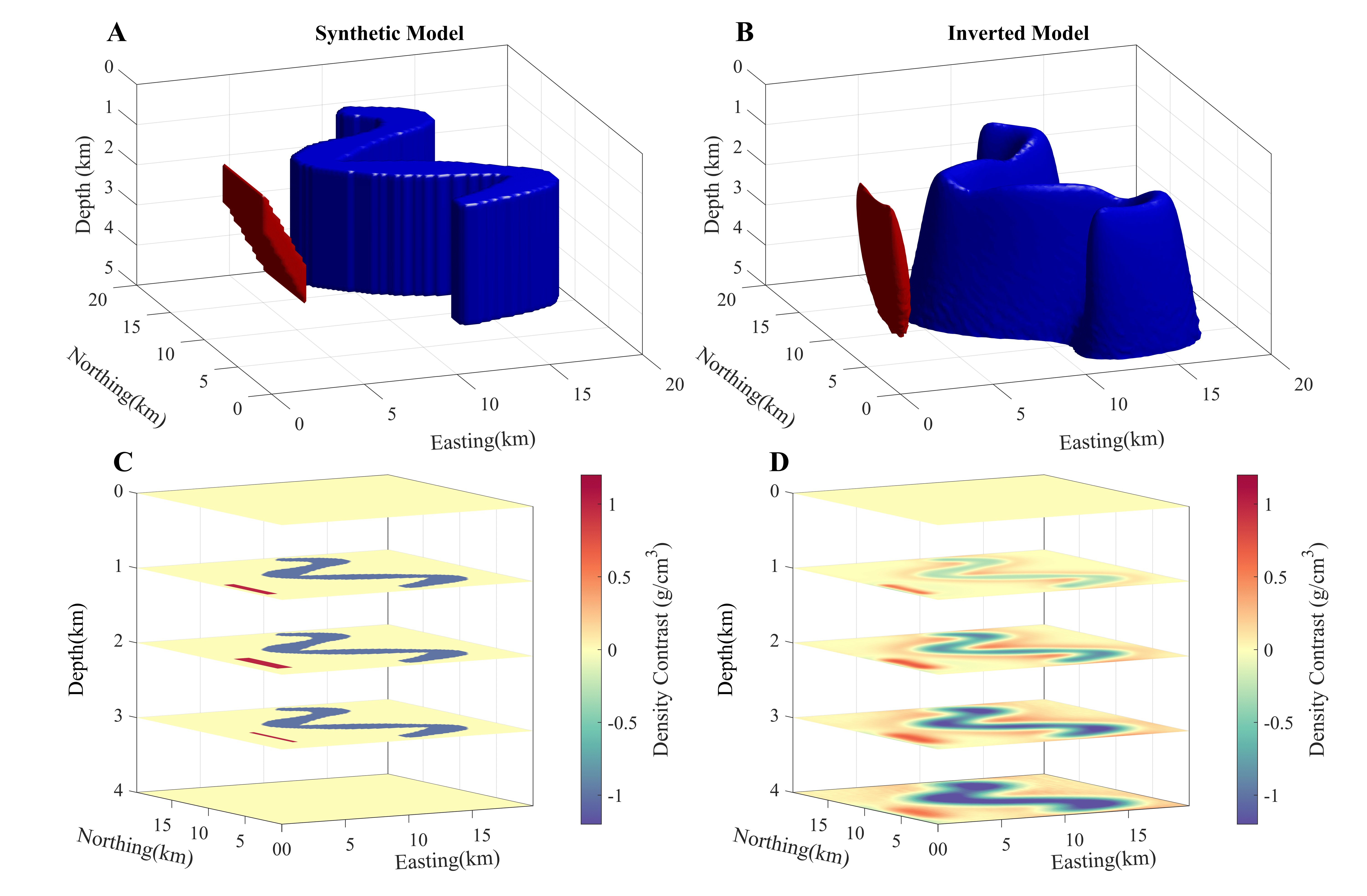}
\caption{{
Complex synthetic model and inversion results containing a curvilinear
tube and a dipping slab. (A) Three-dimensional representation of the
true synthetic model, shown using isosurfaces at density contrasts of
$-0.5$ and $+0.5$~g/cm$^3$. (B) Three-dimensional representation of the
inverted model using the same isosurface levels. (C) Horizontal depth
slices through the true model. (D) Corresponding depth slices through the
recovered model. The complex synthetic model contains geometrically
distinct bodies with curved boundaries, variable width, and dipping
geometry, providing a more challenging test of the proposed data space
inversion framework.
}}
\label{fig:complex_synthetic}
\end{figure}

Figure~\ref{fig:complex_synthetic} compares the true and recovered models.
The three-dimensional isosurface representations show that the inversion
recovers the two geometrically distinct anomalies, while the depth slices
demonstrate the spatial distribution of the recovered density contrasts.
Although some smoothing and distortion are introduced in the recovered
models, particularly near the boundaries of the curvilinear and dipping
structures, the principal geometry and spatial separation of the two
bodies are preserved. 
The inversion successfully recovers the overall geometry and spatial
extent of the anomalies, although some smoothing of density amplitudes
and depth boundaries is observed. This is expected due to the
non-uniqueness of gravity inversion and the adopted regularization.
Improved amplitude and depth resolution could potentially be achieved
by incorporating additional geological constraints or complementary
geophysical data.The present data space formulation is deterministic and does not
generate an ensemble of posterior models. Each inversion produces a
single final density model through iterative updates, and the
observed--calculated gravity difference shown in the results therefore
corresponds to this final model. Uncertainty quantification in terms of
posterior variance or probability maps would require an additional
probabilistic or ensemble-based formulation, which is outside the scope
of the present study.

This more challenging synthetic example demonstrates that the proposed
data space inversion framework is not restricted to simple
axis-aligned rectangular bodies and can recover subsurface structures
with curved boundaries, variable lateral width, and dipping geometry.

\subsection{Tangarparha, Odisha, India – Chromite Ore Body Exploration}

\begin{figure}[h!]
\centering
\includegraphics[width=0.8\textwidth]{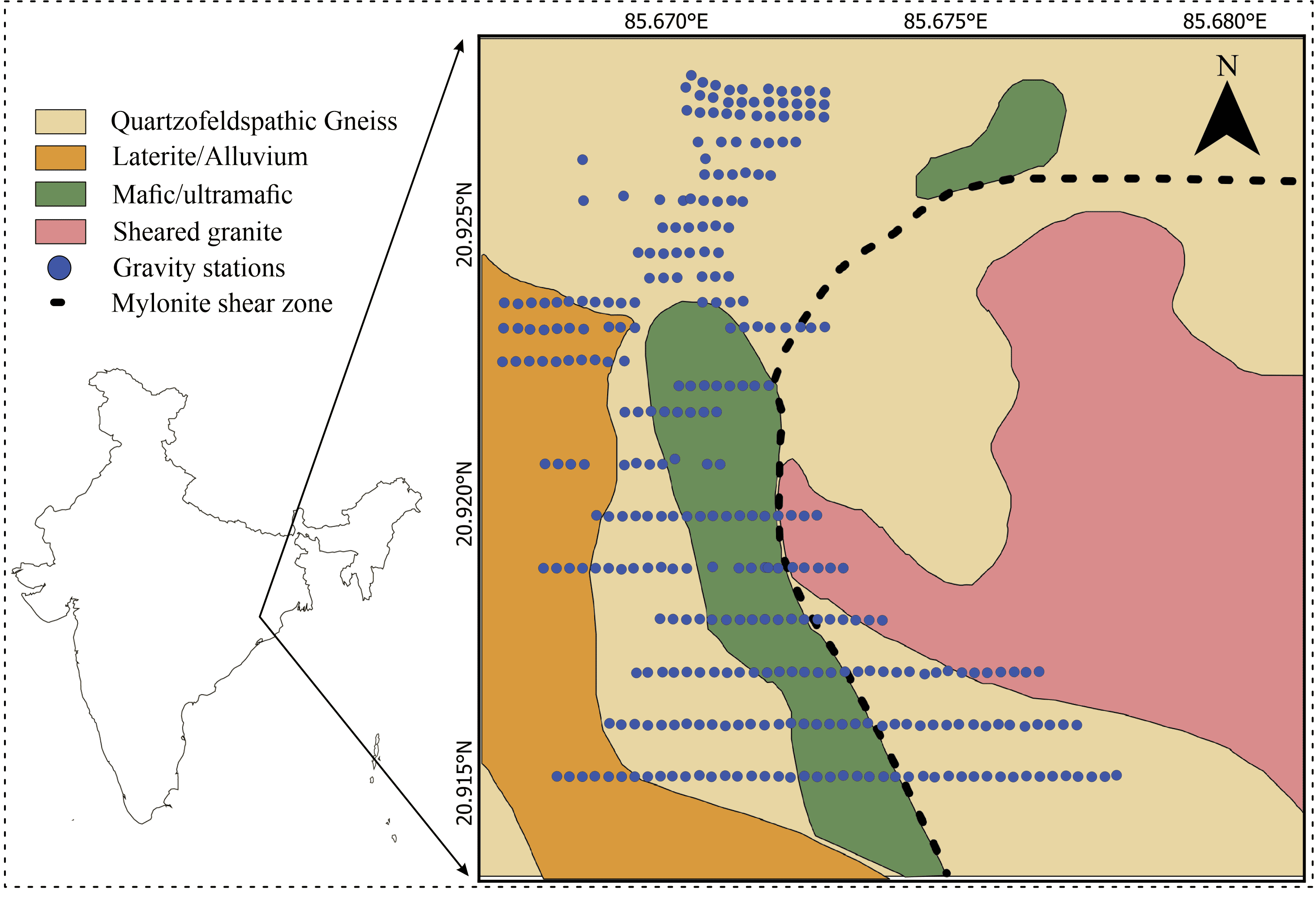}
\caption{Geological map of the Tangarparha study area, Odisha, India, illustrating the distribution of lithological formations such as sheared granite, quartzofeldspathic gneiss, and mafic/ultramafic rocks and key structural features.}
\label{fig:Figure8}
\end{figure}

To assess the effectiveness of the proposed Julia-based 3D gravity inversion framework, we applied it to real-world field data from Tangarparha, Odisha, India—a site previously investigated by \cite{mohanty2011integrated} for chromite exploration. Their study integrated geological and geophysical observations to characterize a $5\,\text{km}^2$
 area comprising several lithological units, including sheared granite, quartzofeldspathic gneiss, and mafic/ultramafic rocks (Figure~\ref{fig:Figure8}). Gravity anomalies within this region are attributed to tectonically deformed and metamorphosed ultramafic bodies, which act as key indicators of subsurface geological structures.
The resulting residual Bouguer gravity anomaly was used as the input
for the 3D inversion.

Gravity measurements were acquired using a W. Sodin gravimeter with an
instrument constant of 0.1082~mGal/division. The fine-counter reading
resolution corresponds to approximately 0.01082~mGal (10.82~$\mu$Gal).

\begin{figure}[h!]
\centering
\includegraphics[width=1\textwidth]{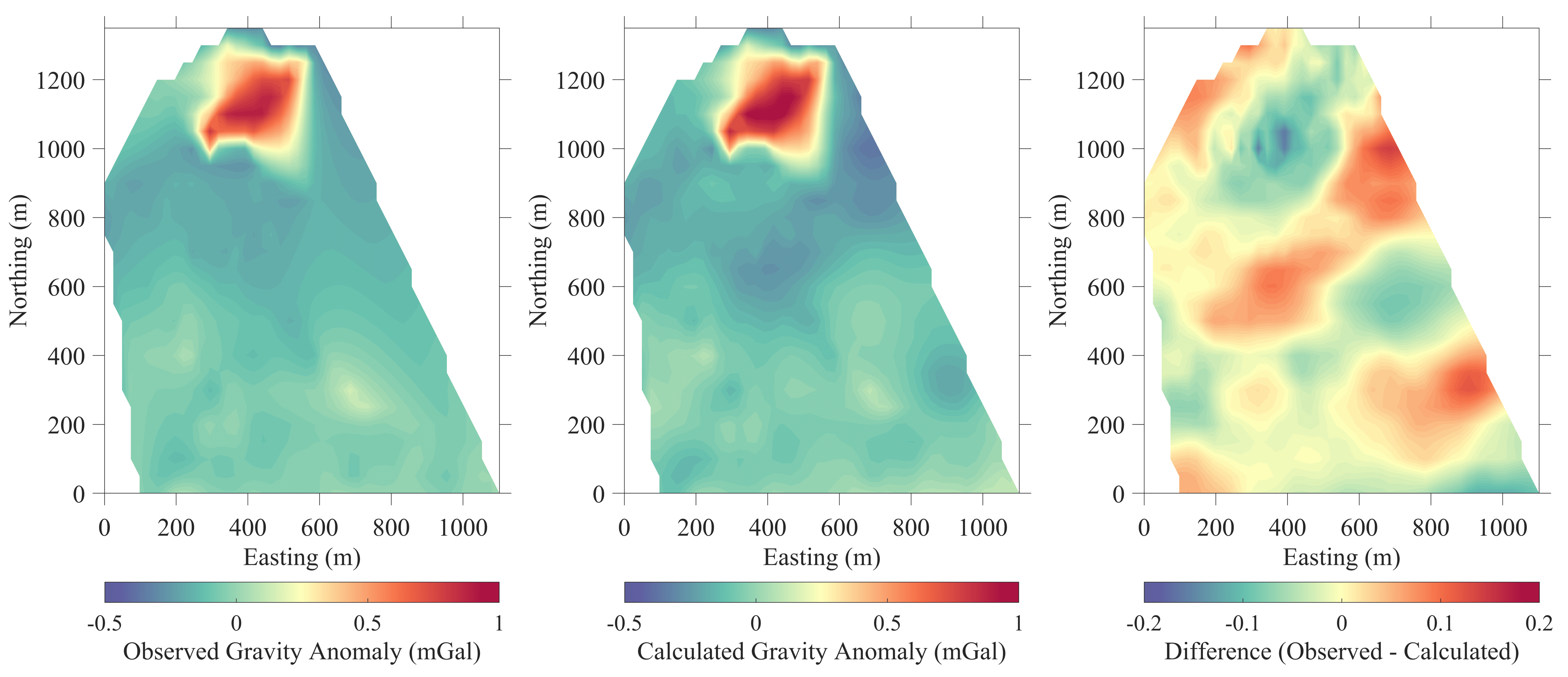}
\caption{Comparison of residual gravity anomaly maps: observed residual gravity anomaly (left), gravity anomaly calculated from the final 3D inversion model (middle), and the difference map between observed and calculated anomalies (right). The strong similarity in spatial patterns and amplitudes between the observed and calculated fields, together with the low-magnitude residuals, demonstrates the effectiveness of the inversion in resolving subsurface density variations.}
\label{fig:Figure9}
\end{figure}

The reliability of our inversion framework was evaluated by comparing the computed gravity response with the observed field data (Figure~\ref{fig:Figure9}). The close agreement between the two datasets indicates that the inversion successfully captures the major subsurface density variations. These findings are consistent with previous geological interpretations suggesting that the region is dominated by quartzofeldspathic gneiss and sheared granite, which serve as host rocks for intrusive mafic/ultramafic bodies. The intrusions include hypabyssal dykes and coarse-grained plutonic units exhibiting cumulate textures, indicative of crystallization within a magma chamber. Chromite mineralization in the area is attributed to magma mixing and crustal assimilation processes within these ultramafic units.

\begin{figure}[h!]
    \centering
    \includegraphics[width=1\textwidth]{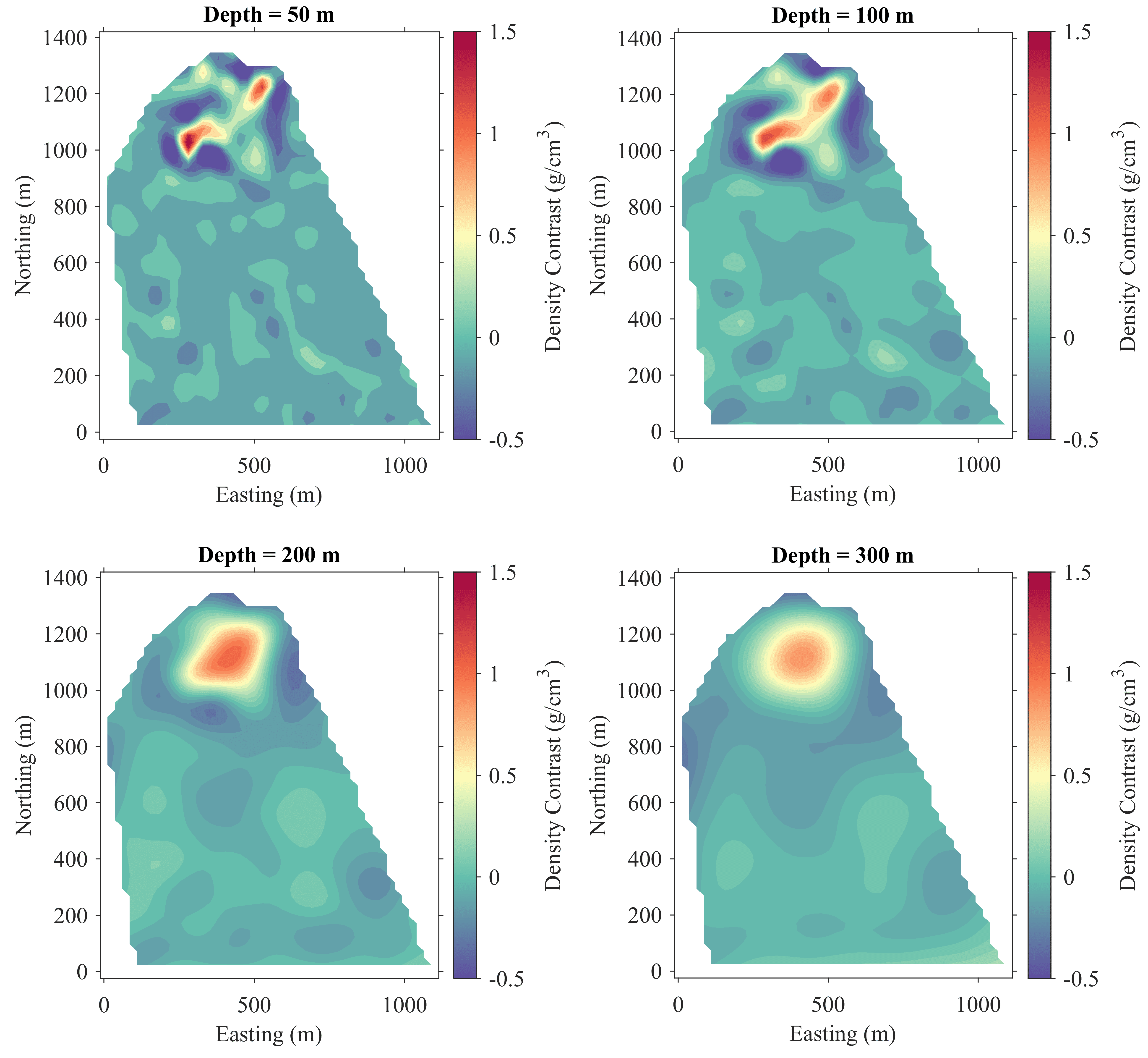}
    \caption{Horizontal depth slices of subsurface density contrast at
    50--300~m depth from the 3D gravity inversion of the Tangarparha field
    data. High-density zones suggest mafic/ultramafic intrusions related to
    chromite mineralization.}
    \label{fig:depth_wise_tangarparha}
\end{figure}

To investigate the internal structure of the subsurface in greater detail, 
horizontal depth slices were extracted from the recovered density model  (Figure~\ref{fig:depth_wise_tangarparha}). These slices, spanning depths of 50–300~m, reveal a persistent high-density anomaly that extends vertically through multiple levels. This continuity supports the presence of a substantial mafic/ultramafic intrusion. Moreover, the variation in the anomaly’s shape and magnitude with depth provides important constraints on the geometry, thickness, and orientation of the intrusive body, further supporting its association with chromite mineralization in the region.

The recovered density contrast reaches a maximum value of approximately
1.5~g/cm$^3$. Using the reference density of 2.67~g/cm$^3$, this corresponds
to an effective bulk density of approximately 4.17~g/cm$^3$. This value is
lower than the density of pure chromite, as the inversion estimates an
effective density contrast over finite model cells rather than the density
of individual chromite grains. The recovered value is therefore interpreted
as representing a chromite-bearing mafic/ultramafic volume rather than pure
chromite, consistent with the geological setting of the Tangarparha area.

\begin{figure}[h!]
\centering
\includegraphics[width=1\textwidth]{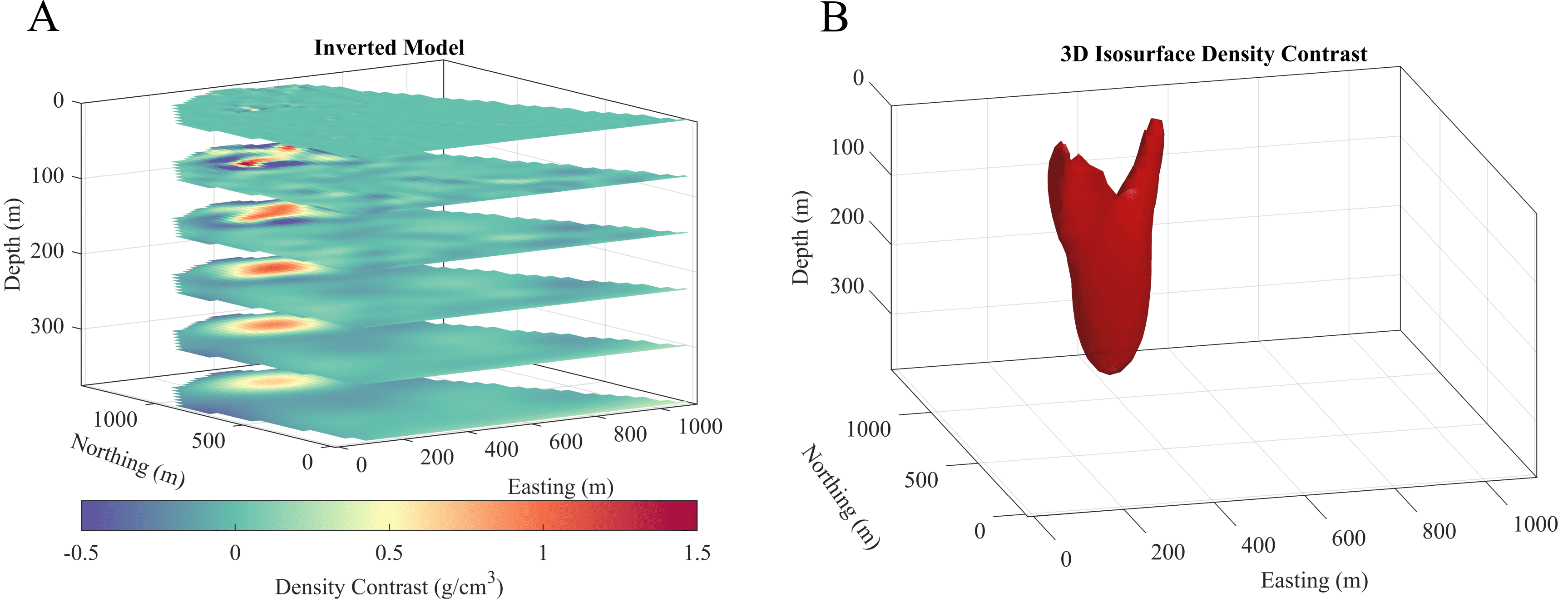} 
\caption{Combined visualization of the 3D inversion results for the Tangarparha study area. The left panel shows the inverted density slice revealing depth-wise density variations and high-density ultramafic zones. The right panel shows the 3D isosurface modelat a density-contrast threshold of 0.7~g/cm$^3$, illustrating the geometry and continuity of the dense body responsible for the gravity anomaly.}
\label{fig:Figure10_11}
\end{figure}

A combined visualization of the inversion output, including both a density slice and a 3D isosurface representation, is shown in Figure~\ref{fig:Figure10_11}. The slice offers a depth-resolved view of the density contrast distribution, highlighting localized high-density zones corresponding to the ultramafic intrusion. The accompanying 3D isosurface model delineates the geometry, lateral continuity, and spatial extent of the dense body. Together, these visualizations provide a comprehensive interpretation of both the internal density structure and the overall morphology responsible for the observed gravity anomalies.


\subsection{Bharatpur District, Rajasthan, India – BIF and Subsurface Density Mapping}

The second field application is located in Bharatpur District, Rajasthan,
India, and focuses on recovering subsurface density variations associated
with BIF and other geological formations.
The area is characterized by a pronounced gravity high, indicating the presence of dense subsurface formations. Field surveys and laboratory analysis of rock samples reveal a wide range of lithologies, with banded iron formations (BIF) exhibiting the highest measured density ($\sim$3.746 g/cm\textsuperscript{3}) and sandstone the lowest ($\sim$2.5 g/cm\textsuperscript{3}). 

The residual Bouguer gravity anomaly obtained after regional
separation was used as the input for the 3D inversion.

Geologically, the area is located in the southeastern part of Bharatpur District and lies along the Great Boundary Fault zone, marking the transition between the Vindhyan Basin and the Delhi Supergroup. The lithological sequence includes formations of the Bhander Group (Vindhyan Supergroup), dominated by sandstones and shales, and older rocks of the Alwar and Ranthambhore Groups (Delhi Supergroup), composed primarily of quartzites and phyllites. In several locations, these units are overlain by Quaternary alluvium. The region is structurally complex, with both regional and local faults contributing to the observed gravity variations and influencing subsurface density distributions.

The gravity survey integrates measurements collected along access roads, cart tracks, canal corridors, and footpaths, ensuring broad spatial coverage. Density measurements from field samples confirm considerable variability, reflecting the mixture of sedimentary and metamorphic lithologies within the region. Elevation corrections using differential GPS (DGPS) further refine the dataset, supporting the interpretation of structural controls likely associated with tectonic movements along the Great Boundary Fault. The geological characterization is based on information compiled from \cite{singh2019triangular}. Gravity observations were acquired using a Scintrex CG-5 type relative
gravimeter. The instrument has a reading resolution of 1~$\mu$Gal and a
reported standard field repeatability of less than 5~$\mu$Gal.

\begin{figure}[h!]
\centering
\includegraphics[width=0.8\textwidth]{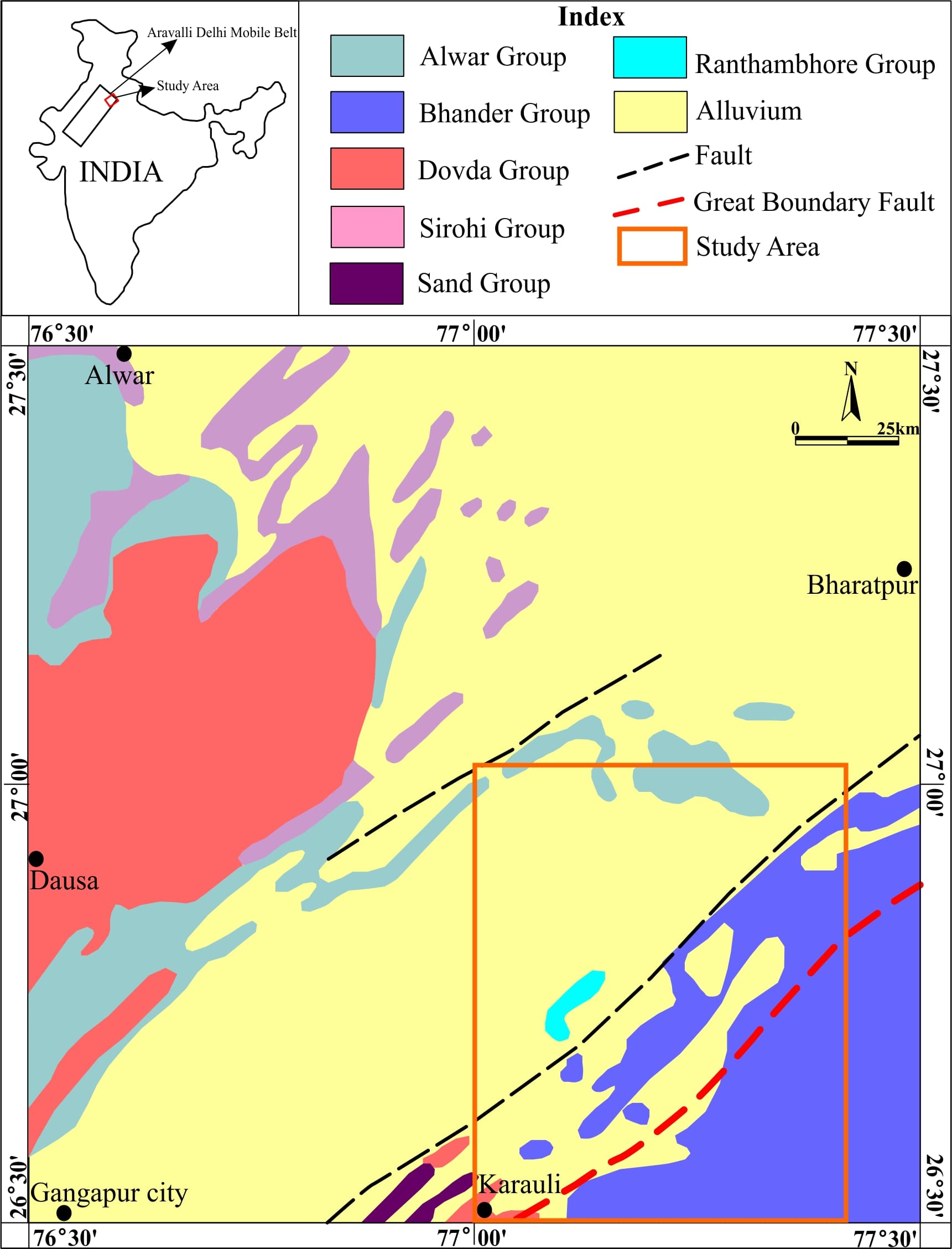} 
\caption{Geological map of the study area located in southeastern Bharatpur District, Rajasthan, India. The map highlights key lithological units, including formations from the Bhander, Alwar, Ranthambhore, Dovda, Sirohi, and Sand Groups, along with Quaternary alluvium. The study area lies along the Great Boundary Fault zone, marking the tectonic transition between the Vindhyan Basin and the Delhi Supergroup. Structural features such as regional faults and the prominent Great Boundary Fault are shown, providing crucial context for interpreting gravity anomalies and subsurface heterogeneity in the region.}
\label{fig:Study_are_map}
\end{figure}

\begin{figure}[h!]
  \centering
  \includegraphics[width=1\textwidth]{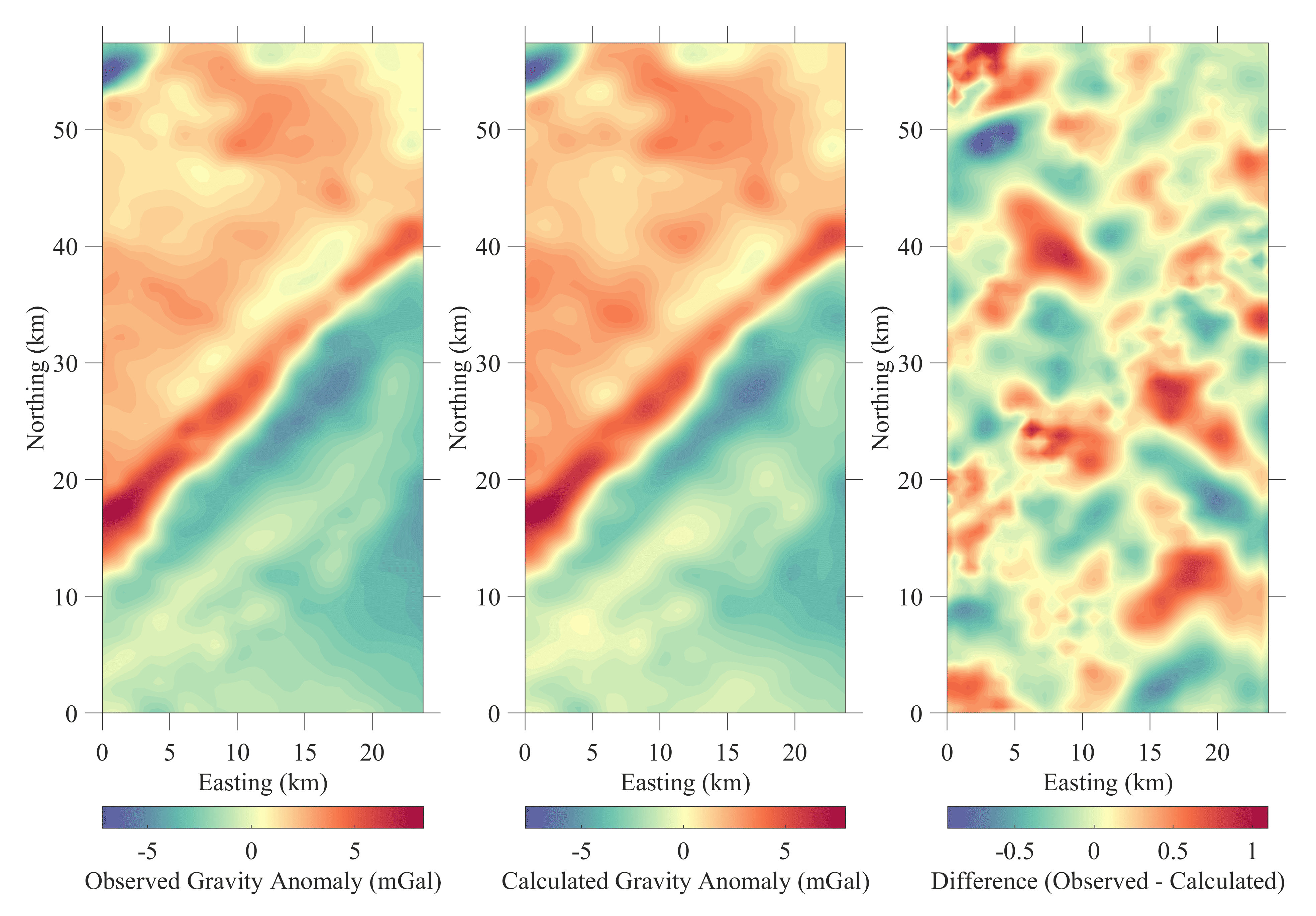} 
  \caption{Comparison of residual gravity anomaly maps: observed residual gravity anomaly (left), gravity anomaly calculated from the final 3D inversion model (middle), and the difference map between observed and calculated anomalies (right). The strong similarity in spatial patterns and amplitudes between the observed and calculated fields, together with the low-magnitude residuals, demonstrates the effectiveness of the inversion in resolving subsurface density variations.}
  \label{fig:field_2}
\end{figure}

Figure~\ref{fig:Study_are_map} presents the geological map of the study region, outlining the major lithological units and structural elements. Figure~\ref{fig:field_2} shows the comparison between observed and computed residual gravity anomalies, illustrating the consistency of the inversion model with field measurements.

\begin{figure}[h!]
  \centering
  \includegraphics[width=1\textwidth]{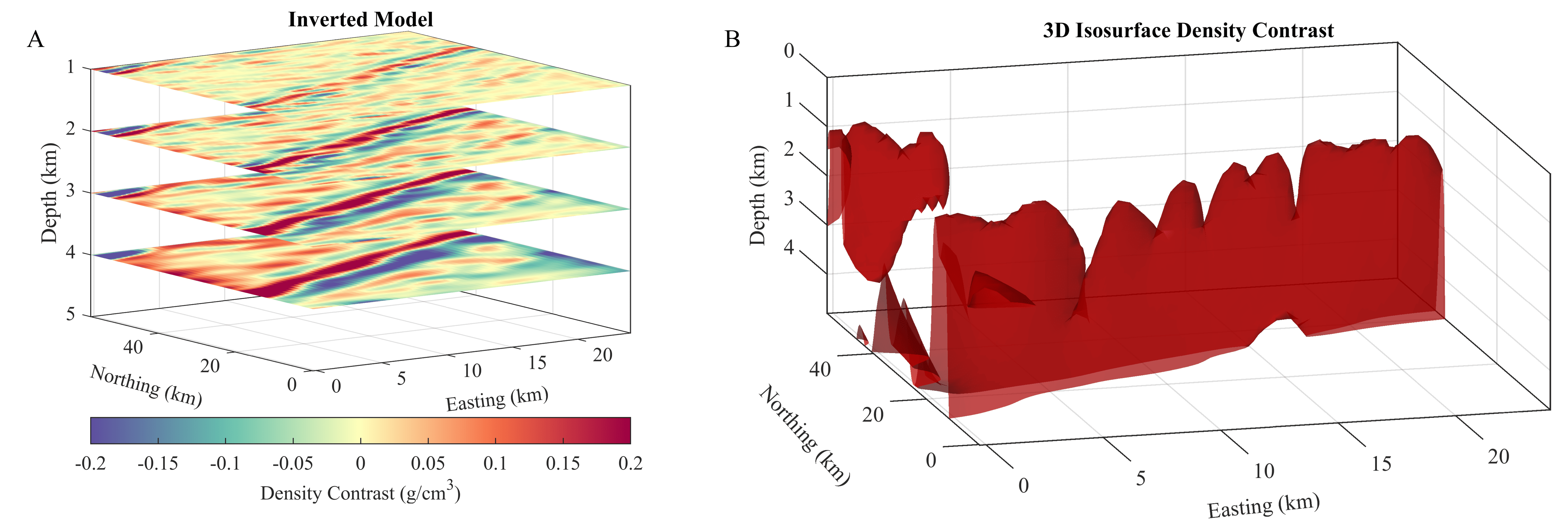} 
 \caption{Combined visualization of the 3D gravity inversion model showing (left) vertical slice plots of the recovered density contrast and (right) a 3D isosurface highlighting high-density formations. The slices depict lateral and vertical variations in density, whereas the isosurface at a density-contrast threshold of 0.15~g/cm$^3$ reveals the spatial geometry and continuity of dense bodies such as banded iron formations (BIF). Together, they provide a comprehensive understanding of the subsurface structure and its relation to tectonic features associated with the Great Boundary Fault zone.}
  \label{fig:combined_slice_iso_2}
\end{figure}

The recovered 3D density structure is visualized through a combined slice and isosurface representation (Figure~\ref{fig:combined_slice_iso_2}). The slice sections reveal pronounced lateral and vertical heterogeneity, with high-density zones—shown in red—corresponding to BIF and other dense geological units that agree with mapped lithologies. The accompanying isosurface visualization captures the three-dimensional geometry, continuity, and orientation of these dense bodies. The structures exhibit eastward-plunging trends aligned with the regional tectonic framework of the Great Boundary Fault. By integrating both slice and volumetric visualizations within a single figure, the model offers an intuitive and comprehensive perspective of the subsurface, thereby strengthening confidence in the geological interpretation.

\begin{figure}[h!]
\centering
\includegraphics[width=0.85\textwidth]{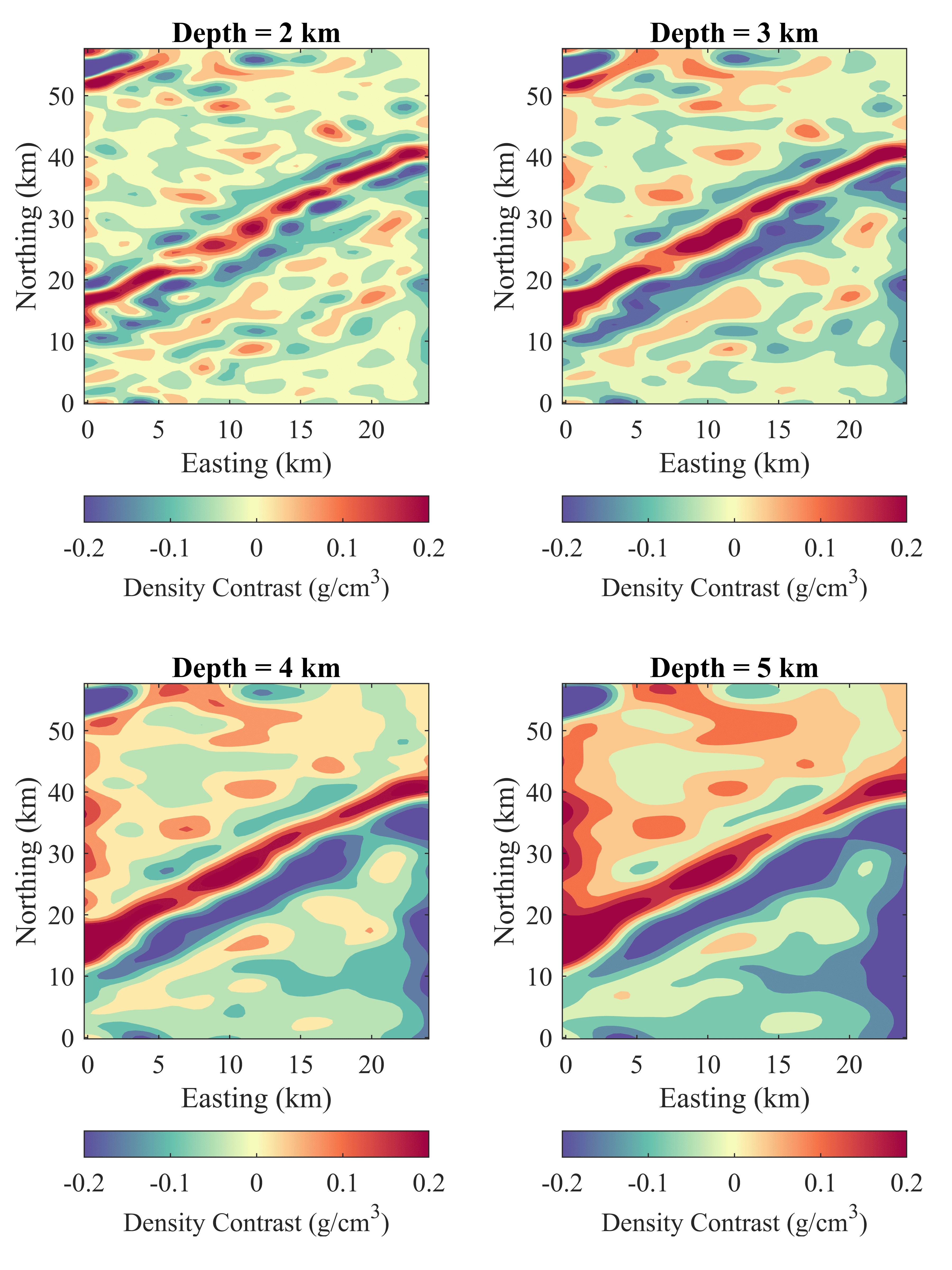} 
\caption{Depth-wise contour maps of subsurface density contrast at 2 km, 3 km, 4 km, and 5 km depth levels derived from 3D gravity inversion. These plots reveal lateral and vertical heterogeneities in the subsurface, highlighting high-density zones interpreted as Banded Iron Formations (BIF) and other dense geological structures. The progressive depth slices aid in visualizing the spatial extent and continuity of subsurface anomalies, supporting tectonic interpretations related to the Great Boundary Fault zone.}
\label{fig:depth_wise_2}
\end{figure}

Finally, Figure~\ref{fig:depth_wise_2} presents depth-wise contour maps from 2 km to 5 km, providing a layered view of density variations at depth and assisting in delineating zones of potential geological significance.

\subsection{Large-Scale Regional Validation}
\label{sec:large-scale-validation}

To further evaluate the scalability and practical applicability of the
proposed multi-GPU sharded solver, the inversion was extended to a
regional field dataset from the Bhukia area of southern Rajasthan,
India. The study region extends approximately from 23$^\circ$30$'$
to 25$^\circ$30$'$N and 73$^\circ$30$'$ to 74$^\circ$45$'$E and covers
approximately 128~km in the east--west direction and 220~km in the
north--south direction, corresponding to an area of about
28,365~km$^2$. The regional data contained 178,797 observations. A single dense sensitivity matrix for
the complete regional problem with 1,803,200
cells would require approximately 1.29~TB of
memory in single precision (Float32), which substantially exceeds the
available memory of the four NVIDIA A100 GPUs. The regional domain was
therefore divided into three adjoining sections, which were inverted
separately using the multi-GPU sharded solver described in
Section~\ref{sec:multigpu-method}. The resulting density models were
then combined to obtain a regional model.

\begin{figure}[t]
\centering
\includegraphics[width=\linewidth]{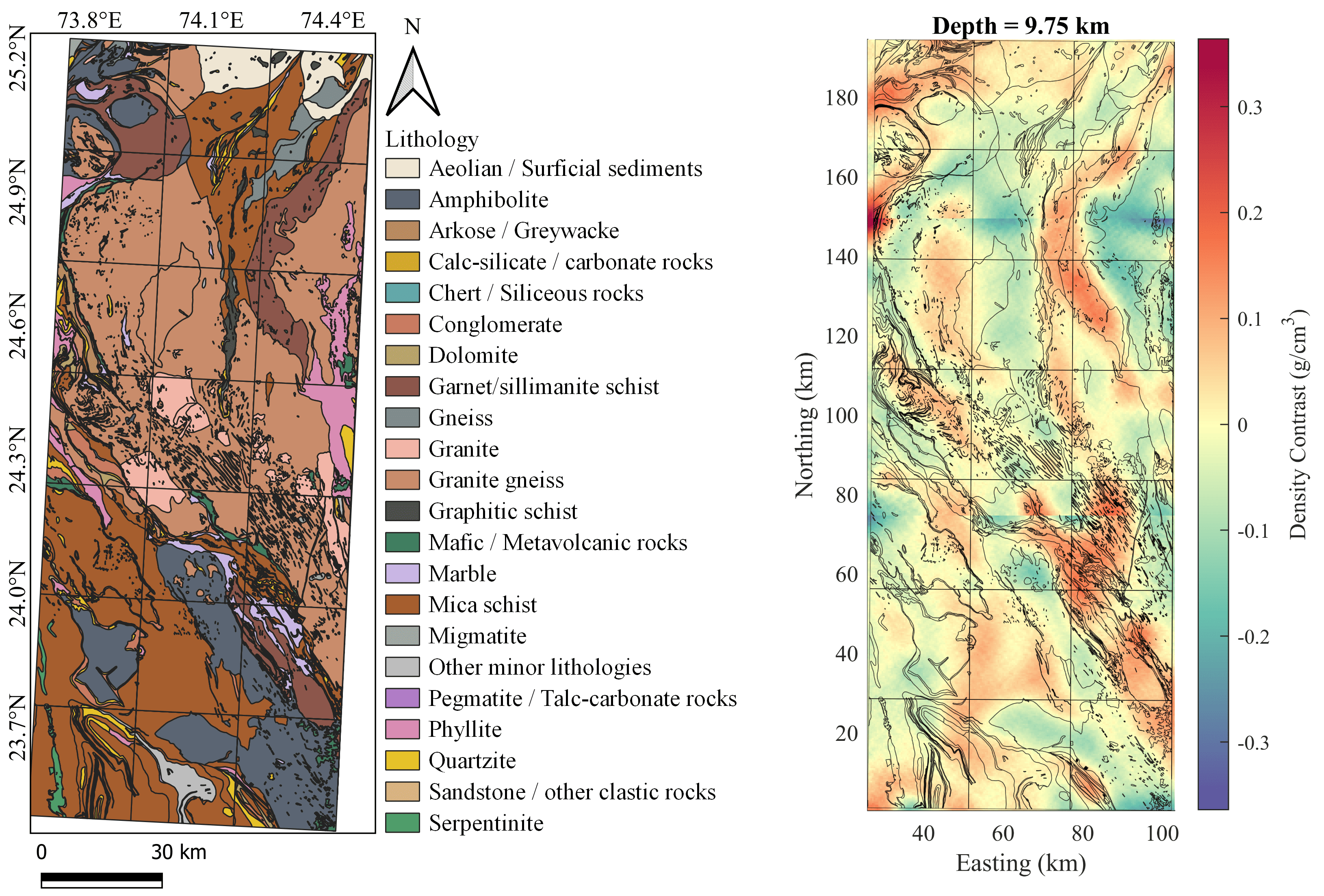}
\caption{Surface geological map of the study area (left) and recovered
density contrast at 9.75~km depth (right) from the combined three-section
multi-GPU inversion. Lithological contacts are shown as overlays on the
density model.}
\label{fig:field-geology-map}
\end{figure}

Because each section was inverted independently, small differences in
the polynomial trend between adjacent sections produced discontinuities
at their boundaries. To reduce these effects, the section models were
combined and a single polynomial trend surface was fitted across the
complete regional domain before visualization. This procedure reduced
the artificial differences between adjoining sections and produced a
more continuous regional density model, as shown in
Fig.~\ref{fig:field-geology-map}.

The computational characteristics of the three sections are summarized
in Table~\ref{tab:scalability}. The first two sections cover
9,472~km$^2$ and contain 602,140 cells and 59,706 observations each.
The third section is slightly smaller, with an area of 9,420.8~km$^2$,
598,920 cells, and 59,385 observations. All three sections were
processed using four NVIDIA A100 GPUs and required approximately
119~s per section, including sensitivity-matrix construction, inversion,
and prediction. The combined processing time for the three independent
runs was 356.87~s, or approximately 6~min. The dense sensitivity
matrices for the individual sections required 143.81, 143.81, and
142.27~GB, respectively, corresponding to approximately 36~GB per GPU
when distributed across four A100 GPUs. The combined size of the three
section matrices was approximately 429.9~GB. This large-scale
experiment demonstrates that the sharded implementation can process
regional-scale problems that cannot be accommodated as a single dense
sensitivity matrix on the available GPU hardware.

\begin{figure*}[t]
\centering
\includegraphics[width=\linewidth]{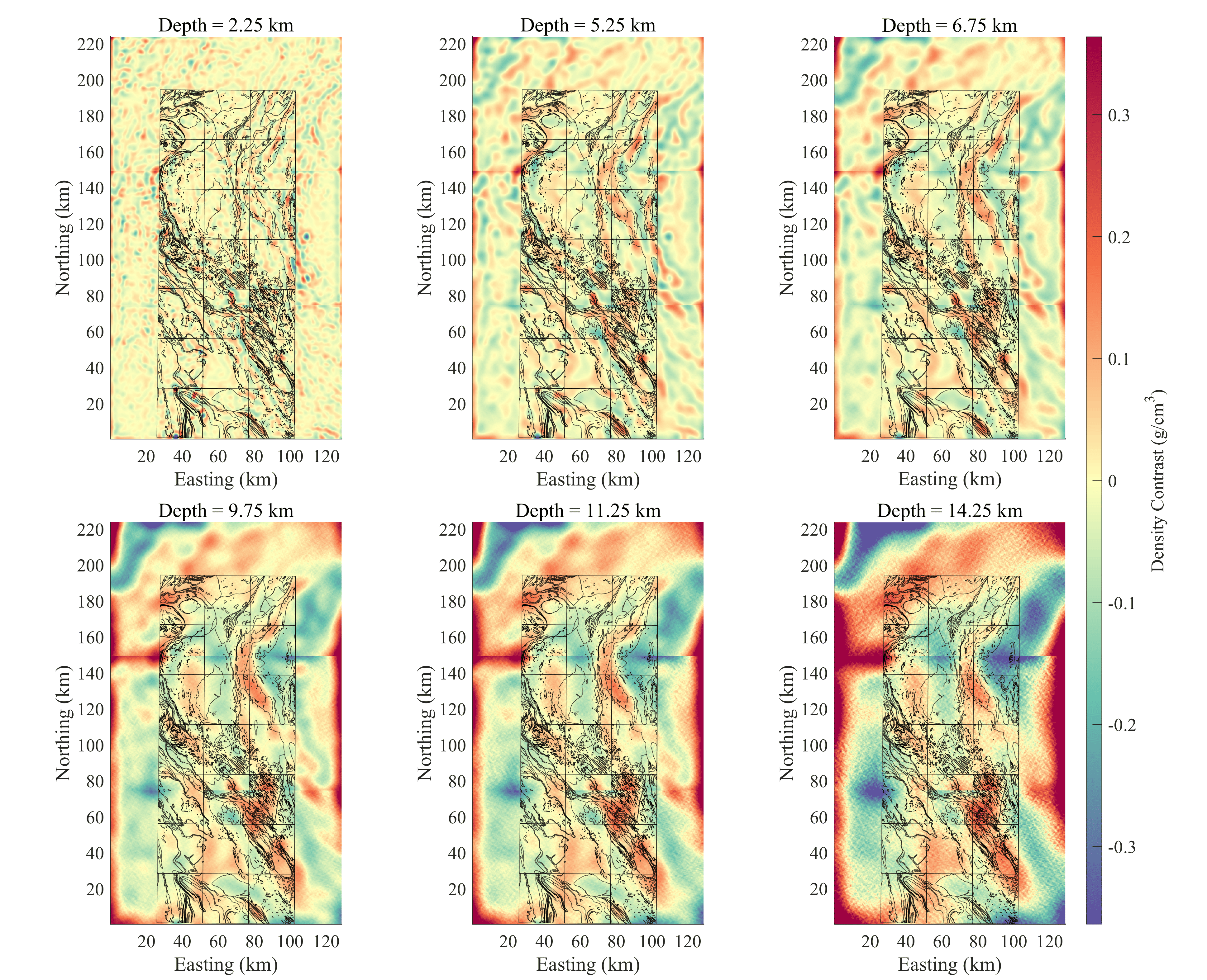}
\caption{Recovered density contrast (color) and surface geological
outlines (black) at depths of 2.25, 5.25, 6.75, 9.75, 11.25, and
14.25~km from the combined regional inversion.}
\label{fig:field-depth-slices}
\end{figure*}

The recovered regional density model was also compared with the surface
geological map to assess the consistency of the larger-scale density
variations with the regional geological framework
(Figs.~\ref{fig:field-geology-map} and
\ref{fig:field-depth-slices}). At shallow depth (2.25~km), the density
model shows relatively heterogeneous and short-wavelength variations,
with limited correspondence to the mapped geological boundaries. At
intermediate depths of approximately 5.25--6.75~km, the recovered
density distribution becomes smoother and several elongated anomalies
develop a predominantly NE-trending pattern that broadly follows the
regional geological fabric.

At 9.75~km depth, several coherent positive and negative density
variations are present across the study area. The stronger positive
density contrasts in the northern and central-southern parts of the
model spatially overlap areas mapped predominantly as
garnet/sillimanite schist. The mica-schist-bearing regions also show
positive density contrasts, although these are generally weaker than
those observed in the garnet/sillimanite-schist domains. Several of the
elongated anomalies follow the regional geological fabric and broadly
coincide with major lithological boundaries. These spatial
relationships suggest that some of the larger-scale density variations
recovered by the inversion are consistent with the regional geological
framework.

At greater depths, between approximately 11.25 and 14.25~km, the
density anomalies become broader and their direct correspondence with
the surface geology decreases. This behavior is consistent with the
decreasing depth resolution of gravity data and the inherent
non-uniqueness of potential-field inversion. Features near the model
margins should also be interpreted cautiously because they may be
affected by the finite observation extent and boundary effects.
Overall, the regional inversion recovers laterally coherent density
structures that broadly follow the regional geological trends,
particularly at intermediate depths, while the deeper model becomes
progressively broader and less directly related to the surface
geological units.

\begin{table}[t]
\centering
\caption{Scale and runtime characteristics of the three regional
sections processed using four NVIDIA A100 GPUs. Dense $\mathbf{G}$ size
is reported in single precision (Float32).}
\label{tab:scalability}
\begin{tabular}{@{}lrrrr@{}}
\toprule
\textbf{Metric} & \textbf{Sec.\ 1} & \textbf{Sec.\ 2} & \textbf{Sec.\ 3} & \textbf{Combined} \\
\midrule
Domain extent (km)
& $128.0 \times 74.0$
& $128.0 \times 74.0$
& $128.0 \times 73.6$
& $128.0 \times 221.6$ \\

Domain area (km$^2$)
& $9{,}472.0$
& $9{,}472.0$
& $9{,}420.8$
& $28{,}364.8$ \\

Mesh ($n_x \times n_y \times n_z$)
& $322\times187\times10$
& $322\times187\times10$
& $322\times186\times10$
& --- \\

Total cells
& $602{,}140$
& $602{,}140$
& $598{,}920$
& $1{,}803{,}200$ \\

Observations $n_{\mathrm{obs}}$
& $59{,}706$
& $59{,}706$
& $59{,}385$
& $178{,}797$ \\

GPUs used
& 4 & 4 & 4 & 12 \\

Dense $\mathbf{G}$ size (GB)
& 143.81 & 143.81 & 142.27 & 429.88 \\

Wall-clock time (s)
& 118.61 & 119.16 & 119.10 & 356.87 \\

RMS data misfit (mGal)
& 0.662 & 0.706 & 0.776 & --- \\
\bottomrule
\end{tabular}
\end{table}

\section{Discussion}

The results demonstrate that the proposed data space formulation can
efficiently solve large-scale 3D gravity inversion problems while
maintaining comparable data misfit to the CPU implementation. The
synthetic experiments show that the principal anomaly geometries are
successfully recovered, while the noise-sensitivity analysis indicates
that increasing observational noise progressively reduces the
resolution of weaker and deeper structures. Nevertheless, the main
anomalies remain identifiable at the tested noise levels.

The computational benchmarks demonstrate that GPU acceleration becomes
particularly beneficial as the model size increases. Reformulating the
problem from the $M\times M$ model-space system to the $N\times N$
data space system reduces the size of the dense system involved in the
inversion, while the multi-GPU implementation distributes the memory
requirement across devices. The Bhukia regional example further
demonstrates the ability of the framework to handle large-scale
problems that cannot be accommodated as a single dense sensitivity
matrix on the available GPU memory.

The field applications show that the recovered density structures are broadly
consistent with the available geological information. For the Tangarparha
example, the recovered density contrast reaches approximately
$1.5~\mathrm{g/cm^3}$, corresponding to an effective bulk density of about
$4.17~\mathrm{g/cm^3}$ when referenced to a background density of
$2.67~\mathrm{g/cm^3}$. This value should be interpreted as an effective
density of the unresolved chromite-bearing mafic/ultramafic volume rather
than the density of pure chromite. For Bharatpur, the recovered high-density
regions are consistent with the independently reported high-density BIF
samples, with measured densities of approximately $3.746~\mathrm{g/cm^3}$.
These comparisons support the geological interpretation of the recovered
anomalies, while the inherent non-uniqueness of gravity inversion means that
the recovered densities should be regarded as plausible effective values
rather than unique petrophysical estimates.

The present study uses ground-based gravity observations for all
real-data applications. Although the computational framework is
demonstrated using NVIDIA CUDA GPUs, its backend-agnostic formulation
provides a basis for extension to other computing architectures.
Similarly, the data space formulation is not restricted to land
gravity and could be extended to larger airborne or satellite-derived
gravity datasets. Such applications would require appropriate
treatment of their different sampling, altitude, spatial-resolution,
and error characteristics and are therefore left for future work.

\FloatBarrier
\section{Conclusion}

This study presents a high-performance three-dimensional gravity inversion framework implemented in Julia to address key challenges in geophysical modeling, including computational inefficiency, non-uniqueness, and the ill-posed nature of gravity inversion. By employing a data space inversion strategy, the framework reduces the dimensionality of the inverse problem, resulting in substantial improvements in computational speed and memory efficiency while preserving inversion accuracy. The incorporation of GPU acceleration further enhances performance, yielding one to two orders of magnitude speedup for large-scale models with up to approximately 3.3 million cells compared to conventional CPU-based implementations. This capability enables practical high-resolution subsurface imaging for applications such as resource exploration and tectonic studies.

The effectiveness of the proposed framework was demonstrated using both synthetic and field datasets. Synthetic experiments showed that the method can reliably reconstruct geometrically complex subsurface structures, including vertical and dipping dykes, while exhibiting expected smoothing near anomaly boundaries due to the inherent non-uniqueness of gravity data. Field applications from Odisha and Rajasthan, India, further confirmed the robustness of the approach, with inversion results showing strong agreement with known geological features such as ultramafic intrusions and banded iron formations. 
The large-scale Bhukia example further demonstrated the computational
scalability of the data space formulation for large gravity inversion
with millions of model cells. The close correspondence between observed and computed gravity responses highlights the reliability of the framework for real-world gravity inversion problems.

By leveraging Julia’s high-performance computing ecosystem and GPU parallelism, the proposed framework overcomes traditional computational bottlenecks associated with large-scale 3D gravity inversion. The scalable and portable design of the implementation makes it well suited for modern high-performance computing environments. Future extensions may include joint inversion with complementary geophysical datasets, such as magnetic or seismic data, to further reduce interpretation uncertainty. Overall, this work provides an efficient and flexible tool for large-scale gravity inversion, with clear potential for both academic research and industrial geophysical applications.

\section*{CRediT authorship contribution statement}

\textbf{Nimatullah}: Writing – original draft, Writing – review \& editing, software, methodology, formal analysis, data curation, conceptualization.
\textbf{Pankaj K.\ Mishra}: Writing – review \& editing, software, funding acquisition, conceptualization, supervision. 
\textbf{Jochen Kamm}:  Writing – review \& editing, funding acquisition. 
\textbf{Anand Singh}: Writing – review \& editing, software, funding acquisition, conceptualization, supervision.

\subsection*{Software availability}

Name of the code/library: GravityInversionGPU.jl: A backend-agnostic Julia framework for three-dimensional gravity modeling and inversion

Contact: Nimatullah (24D0455@iitb.ac.in)

Hardware requirements: GPU acceleration experimentally validated on NVIDIA
CUDA/A100 GPUs. The framework provides backend interfaces for CUDA, AMDGPU,
Metal, and oneAPI.

Program language: Julia

Software required: Julia (version $\geq$ 1.9); GPU backend packages (CUDA.jl, Metal.jl, AMDGPU.jl, oneAPI.jl)

Program size: Approximately 0.5~MB (excluding optional dependencies)

Availability of the source code:
The source code is available on Zenodo (\url{https://doi.org/10.5281/zenodo.22171658}) and GitHub (\url{https://github.com/naimat04/GravityInversionGPU.jl}).

\section*{Acknowledgments}

This work was carried out as part of the Ph.D. research of Nimatullah at the Indian Institute of Technology Bombay. The research was financially supported by the Indian Institute of Technology Bombay and the Research Council of Finland under the High-Performance Computing 2023 Project (Grant No. 35926). The authors also acknowledge the Geological Survey of Finland for institutional support and collaboration. Computational resources and research facilities were provided by CSC – IT Center for Science, Finland, and by the Indian Institute of Technology Bombay, both of which are gratefully acknowledged.

\section*{Declaration of generative AI and AI-assisted technologies in the manuscript preparation process}

During the preparation of this manuscript, the authors used ChatGPT (OpenAI) and Grammarly to assist with language refinement, improving clarity and readability, and organizing the presentation of the content. The use of these tools was limited to editorial and linguistic support and did not influence the scientific analysis, interpretation of results, or conclusions of the study. All content generated or modified with the assistance of these tools was carefully reviewed, verified, and edited by the authors, who take full responsibility for the accuracy, originality, and integrity of the published work.

\bibliographystyle{cas-model2-names}
\bibliography{references} 
\end{document}